\documentclass[aps,prl,twocolumn,showpacs,groupedaddress,superscriptaddress,footinbib,longbibliography]{revtex4-2}

\usepackage{mathrsfs}
\usepackage{natbib,hyperref}
\setcitestyle{square,sort&compress,comma,numbers}
\hypersetup{colorlinks=true, citecolor=blue, urlcolor=blue, linkcolor=blue}
\usepackage[english]{babel}
\usepackage[utf8]{inputenc}
\usepackage{graphicx}
\usepackage[caption=false]{subfig}
\usepackage{amsmath}
\usepackage{amsfonts}
\usepackage{amssymb}
\usepackage{color,soul}
\setulcolor{red}
\usepackage{easyReview}
\usepackage{xcolor}
\usepackage{float}
\usepackage{pgfplots}
\usepackage[normalem]{ulem}
\newcommand\diff{\mathrm{d}}
\renewcommand{\vec}[1]{\boldsymbol{#1}}

\renewcommand{\theta}{\vartheta}
\begin{document}

\title{Hydroelastic Scattering and Trapping of Microswimmers}
\author{Sagnik Garai}
\affiliation{Max Planck Institute for the Physics of Complex Systems, N\"othnitzer Stra{\ss}e 38,
01187 Dresden, Germany}
\author{Ursy Makanga}
\affiliation{Max Planck Institute for the Physics of Complex Systems, N\"othnitzer Stra{\ss}e 38,
01187 Dresden, Germany}
\author{Akhil Varma}
\affiliation{Max Planck Institute for the Physics of Complex Systems, N\"othnitzer Stra{\ss}e 38,
01187 Dresden, Germany}
\author{Christina Kurzthaler}
\email{ckurzthaler@pks.mpg.de}
\affiliation{Max Planck Institute for the Physics of Complex Systems, N\"othnitzer Stra{\ss}e 38,
01187 Dresden, Germany}
\affiliation{Center for Systems Biology Dresden, Pfotenhauerstraße 108, 01307 Dresden, Germany}
\affiliation{Cluster of Excellence, Physics of Life, TU Dresden, Arnoldstraße 18, 01062 Dresden, Germany}

\begin{abstract} 
Deformable boundaries are omnipresent in the habitats of swimming microorganisms, leading to intricate hydroelastic couplings. Employing a perturbation theory, valid for small deformations, we study the swimming dynamics of pushers and pullers near instantaneously deforming boundaries, endowed with a bending rigidity and surface tension. Our results reveal that pushers can both reorient away from the boundary, leading to overall hydroelastic scattering, or become trapped by the boundary, akin to the enhanced trapping found for pullers. These findings demonstrate that the complex hydroelastic interactions can generate behaviors that are in striking contrast to swimming near planar walls. 
\end{abstract} 

\maketitle
Swimming microorganisms represent fascinating exemplars of non-equilibrium systems and their dynamics have been widely studied in the realm of fluid mechanics and statistical physics~\cite{Lauga:2009,Marchetti:2013, Bechinger:2016, Lauga:2020, Gompper:2020, Kurzthaler:2023}. Owing to their small size and typical swimming velocity, the effects of fluid inertia are negligible compared to viscous ones, thus requiring cells to perform non-reciprocal swimming strokes to propel through their low-Reynolds-number fluid environments~\cite{Purcell:1977}. Unraveling their swimming behavior in their natural habitats is paramount for gaining new insights into the formation of bacterial colonies and biofilms~\cite{Hartmann:2019, Hallatschek:2023}, the complexities of disease spreading bacteria~\cite{Otte:2021}, and sperm motility in the reproductive tract~\cite{Suarez:2006}, to name a few. They further lay the foundation for advancing technology through the design of novel cargo-carriers~\cite{Golestanian:2019, Volpe:2024, wuMedicalMicroNanorobots2020, nelsonMicrorobotsMinimallyInvasive2010}. 

Many microorganisms live in confined environments, where long-ranged hydrodynamic interactions with nearby boundaries can strongly affect their swimming behaviors, leading to various interesting physical phenomena~\cite{Bechinger:2016}. Seminal experiments revealed a change from clockwise~\cite{Lauga:2006} to counterclockwise circular motion~\cite{diLeonardo:2011} of {\it Escherichia coli} bacteria by varying the boundary from a rigid no-slip wall to an air-water interface. These dynamics have been predicted to change for spherical microswimmers and phoretic colloids, which can scatter, perform oscillatory dynamics, or hover near the wall, depending on the propulsion modes and their initial orientation~\cite{Ishimoto2013, Li:2014, Uspal:2014, Kuron:2019, Desai:2021}. Introducing a boundary slip~\cite{Poddar:2020} or a non-planar boundary shape~\cite{Christina:2021, Ishimoto2023}, can further strongly modify the dynamics, demonstrating that details of the hydrodynamic interactions are crucial for active systems. 

In contrast to rigid walls, many biological surfaces~\cite{Helfrich:1973,Brown:2008,Vutukuri:2020,Takatori:2020,Fessler:2024,Azadbakht:2024}, such as membranes, are soft and can deform in response to hydrodynamic stresses. The inherent non-linearity due to the coupling of the flow and the deformation leads to unexpected physics~\cite{Trouilloud:2008, Shaik:2017, Ledesma:2013, Takatori:2020, Azadbakht:2024, Fessler:2024}, while rendering these problems challenging to handle analytically and numerically.  Instrumental insights come from the context of externally driven spheres, where the hydroelastic coupling induces a lift force, pushing the sphere away from the surface -- a phenomenon not possible near planar walls due to the time-reversibility of Stokes flows~\cite{Skotheim:2004,Bureau:2023,Rallabandi:2024, Davies:2018}. Theoretical progress was made possible by employing the Lorentz reciprocal theorem~\cite{lorentz1897general}, providing insights for particle lift near fluid-fluid interfaces~\cite{Lee_Chadwick_Leal_1979,Berdan:1982} and compliant surfaces~\cite{Rallabandi:2018,Davies:2018}. In contrast to their driven passive counterparts, swimming microorganisms are force- and torque-free, thus producing different flow signatures~\cite{Drescher:2010, Drescher:2011}, which are expected to fundamentally change the hydroelastic interaction. Recent work discovered the emergence of directed motion of agents with reciprocal swim strokes near deformable boundaries~\cite{Trouilloud:2008, Shaik:2017}, their enhanced swim speed through deformable channels~\cite{Ledesma:2013}, and their swimming velocities at the time scale of the deformation~\cite{Daddi-Moussa-ider_PRF:2018,Nambiar:2022}. Yet, a comprehensive understanding of the swimming dynamics near a deformable boundary at larger time scales, their stationary state and its dependence on both material and swimmer properties remains lacking and poses a fundamental ingredient towards unraveling more complex many-body systems~\cite{Vutukuri:2020, Takatori:2020, Azadbakht:2024, Fessler:2024}. 

Here, we study the dynamics of a microswimmer near a deformable boundary, characterized by a bending rigidity and surface tension, by means of a perturbation theory, valid for small deformations. Our main findings demonstrate that the interaction induces scattering of pusher swimmers away from the boundary and enhanced trapping of pullers. For small Föppl-von K{\'a}rm{\'a}n numbers, measuring the relative importance of surface tension to bending rigidity, pushers can reorient towards the boundary, leading to a trapping  state. Most importantly, these features emerge from a hydroelastic interaction, reorienting the active agents, in contrast to the deformation-induced lift of the passive counterpart.

\paragraph{Model.--}We consider a microswimmer in an incompressible fluid of viscosity $\mu$ near an elastic boundary~$\mathcal{S}_M$. It creates a flow described by the spatially-varying fluid velocity field $\boldsymbol{{u}}(\boldsymbol{{r}})$ and pressure field~${p}(\boldsymbol{{r}})$ with $\vec{r}=(\vec{r}_\parallel,z)^T$. Here, $\vec{r}_\parallel$ denotes the components on the undeformed surface $\mathcal{S}_0$ (corresponding to the $xy-$plane) and $z$ is the normal component.
In the low-Reynolds-number limit the flow fields are governed by the quasi-steady Stokes equations~\cite{Kim:2005, Happel:1983, Leal:2007}:
$\boldsymbol{{\nabla}} {p}=\mu{\nabla}^{2}\boldsymbol{{u}}$ and $\boldsymbol{{\nabla}}\cdot \boldsymbol{{u}}=0$. 
As response to its swimming stroke and the presence of surrounding boundaries, 
the microswimmer translates at a velocity~$\boldsymbol{{U}}$ and rotates at an angular velocity~$\boldsymbol{{\Omega}}$. Here, we consider an infinitesimally thin elastic boundary $\mathcal{S}_M$ (unbounded in the other directions), which can deform in response to the viscous stresses $\boldsymbol{{\sigma}}= -p\mathbb{I}+\mu(\boldsymbol{{\nabla} {u}}+\boldsymbol{{\nabla} {u}}^{T})$. Its deformation $z={\delta}(\vec{r}_\parallel)$ obeys the stress balance~\cite{Helfrich:1973, Rallabandi:2018} 
\begin{align}
\left(\kappa{\nabla}_{\parallel}^{4}-\Sigma{\nabla}_{\parallel}^{2}\right){\delta}=\boldsymbol{{\sigma}}:\boldsymbol{n}\boldsymbol{n}|_{\mathcal{S}_M},\label{helfrich}
\end{align}
where $\kappa$ and $\Sigma$ denote the bending rigidity and surface tension, respectively, $\vec{n}$ is the normally-outward pointing unit vector, and $\vec{\nabla}_{\parallel}$ is the in-plane gradient. Note that we have assumed no in-plane deformation of the boundary. The problem set-up is complete after imposing the kinematic and no-slip boundary conditions on~$\mathcal{S}_M$~\cite{Leal:2007}, and assuming that the velocity field vanishes far from the microswimmer (see Supplementary Information~(SI)~\cite{supp}). 

Dimensional analysis reveals that our system has two non-dimensional parameters: (1) a generalized elasto-viscous number, $\epsilon=\mu a^{2}U_{\rm free}/(\kappa+\Sigma a^{2})$, 
measuring the relative importance of the hydrodynamic stress induced by the microswimmer (of size $a$ and free-space swimming speed $U_{\rm free}$) to the elastic resistance of the boundary~\cite{Lee_Chadwick_Leal_1979}, and (2) the Föppl-von K{\'a}rm{\'a}n (FvK) number, $\Gamma = \Sigma a^2/\kappa$, comparing the strength of surface tension with bending rigidity. Typical estimates for {\it E.~coli} near lipid membranes (comprising a majority of biological membranes), are of the order of $\epsilon= \mathcal{O}(10^{-1})-\mathcal{O}(1)$ and $\Gamma = \mathcal{O}(1)-\mathcal{O}(10)$~\cite{Faucon:1989, Vutukuri:2020}. 

\paragraph{Small-deformation limit.--} In this work, we focus on the regime of small elasto-viscous numbers, $\epsilon\ll 1$, i.e. the elastic resistance of the boundary is large compared to the viscous stresses. This allows considering small surface deformations and employing a perturbation ansatz of the form
$\{\delta, \boldsymbol{u},\boldsymbol{U}, \boldsymbol{\Omega},p\} =  \{\delta^{(0)}, \boldsymbol{u}^{(0)},\boldsymbol{U}^{(0)}, \boldsymbol{\Omega}^{(0)},p^{(0)}\}+\epsilon\{\delta^{(1)},\boldsymbol{u}^{(1)},\boldsymbol{U}^{(1)}, \boldsymbol{\Omega}^{(1)}, p^{(1)}\}+\mathcal{O}(\epsilon^{2})$,
where the first set of terms are the solution variables near a planar wall ($\delta^{(0)}=0$) and the second set of terms are the leading-order corrections. Furthermore, employing a Taylor series expansion in $\delta$ about the reference surface $\mathcal{S}_0$ in Eq.~\eqref{helfrich}~\cite{supp}, we find that the leading-order deformation is governed by $\left(\kappa \nabla_{\parallel}^{4}-{\Sigma}\nabla_{\parallel}^{2}\right)\delta^{(1)}(\vec{r}_\parallel)=-p^{(0)}(\vec{r}_\parallel)|_{z=0}$,
permitting a solution in Fourier space ($\vec{r}_\parallel\to\vec{k}$) of the form
\begin{align}
\delta^{(1)}(\boldsymbol{k})=- \dfrac{\left.p^{(0)}(\boldsymbol{k})\right|_{z=0}}{\kappa|\boldsymbol{k}|^{4}+\Sigma |\boldsymbol{k}|^{2}}.\label{eq:def}
\end{align}
A numerical inverse transform then provides $\delta^{(1)}(\boldsymbol{r}_\parallel)$~\cite{supp}. 

\paragraph{Deformation-induced swimming velocities.--}Using the Lorentz reciprocal theorem~\cite{Masoud:2019, Rallabandi:2018} and exploiting that the microswimmer is force- and torque-free, yields a relation for the deformation-induced swimming velocities, $\vec{U}^{(1)}$ and $\vec{\Omega}^{(1)}$ (see SI \cite{supp} for details): 
\begin{align}
\boldsymbol{U}^{(1)}\cdot\boldsymbol{\hat{F}}+\boldsymbol{\Omega}^{(1)}\cdot\boldsymbol{\hat{L}}=\int_{\mathcal{S}_{0}} \, \boldsymbol{n}\cdot \boldsymbol{\hat{\sigma}}\cdot\boldsymbol{u}_{\mathcal{S}_M} \diff S, \label{eq:reciprocal}
\end{align}
where the contribution of the surface deformation enters via an effective slip velocity~\cite{Rallabandi:2018, supp} 
\begin{align}
\vec{u}_{\mathcal{S}_M}=\frac{\partial \delta^{(1)}}{\partial t}\vec{e}_z- \left.\delta^{(1)}\frac{\partial \boldsymbol{u}^{(0)}}{\partial z}\right|_{z=0} - \vec{U}_\parallel^{(0)}\cdot \vec{\nabla}_{\parallel}\delta^{(1)}\vec{e}_z. \label{eq:BC_SM}
\end{align}
It comprises the deformation rate ($\partial_t\delta^{(1)}$) and the contribution due to the deformation shape ($\delta^{(1)}$), reminiscent to that of a rigid, structured surface~\cite{Kurzthaler:2021}. The last term represents the advection of the deformation by the swimming agent, where~$\vec{U}_\parallel^{(0)}$ denotes its zeroth-order velocity parallel to $\mathcal{S}_0$. Equation~\eqref{eq:reciprocal} further depends on the stress tensor~$\boldsymbol{\hat{\sigma}}$ of a passive object of the same shape as the microswimmer (referred to as auxiliary problem), moving near a planar (no-slip) wall at velocities $\boldsymbol{\hat{U}}$ and $\boldsymbol{\hat{\Omega}}$, under the application of an external force~$\boldsymbol{\hat{F}}$ and torque~$\boldsymbol{\hat{L}}$.

Due to the linearity of the Stokes equations, we can generalize Eq.~\eqref{eq:reciprocal} by introducing third-rank tensors $\vec{\hat{\mathcal{T}}}_F$ and $\vec{\hat{\mathcal{T}}}_L$, relating the stresses and velocities of the auxiliary problem via: $\vec{n}\cdot\vec{\hat{\sigma}} =  \vec{n}\cdot\vec{\hat{\mathcal{T}}}\cdot (\boldsymbol{\hat{U}}, \boldsymbol{\hat{\Omega}})^T$ with $\vec{\hat{\mathcal{T}}}=(\vec{\hat{\mathcal{T}}}_F,\vec{\hat{\mathcal{T}}}_L)$~\cite{Elfring:2015}. 
We further note that external forces and torques balance the hydrodynamic forces and torques,
\begin{align}
\begin{pmatrix}
\vec{\hat{F}}\\
\vec{\hat{L}}
\end{pmatrix} = -
\int_{\mathcal{S}_P} \begin{pmatrix}
\vec{n}\cdot\vec{\hat{\sigma}}\\
\vec{R} \times \vec{n}\cdot\vec{\hat{\sigma}}
\end{pmatrix} \diff S = \boldsymbol{\hat{\mathcal{R}}} \cdot \begin{pmatrix}
\hat{\boldsymbol{U}}\\
\hat{\boldsymbol{\Omega}}
\end{pmatrix},
\label{eq:force_torque_passive}
\end{align}
where $\vec{R}$ is the position vector directed from the center to the surface of the swimmer\,($\mathcal{S}_P$) and
\begin{align}
\boldsymbol{\hat{\mathcal{R}}} = -\int_{\mathcal{S}_P} \begin{pmatrix}
\vec{n}\cdot\vec{\hat{\mathcal{T}}}\\
\vec{R} \times \vec{n}\cdot\vec{\hat{\mathcal{T}}}
\end{pmatrix} \diff S,  
\end{align}
is the resistance matrix. Combining Eq.~\eqref{eq:force_torque_passive} and Eq.~\eqref{eq:reciprocal}, the leading-order corrections to the translational and angular velocities take the form
\begin{align}
\begin{pmatrix}
\boldsymbol{U}^{(1)}\\
\boldsymbol{\Omega}^{(1)}
\end{pmatrix} &= \boldsymbol{\hat{\mathcal{R}}}^{-1} \cdot \int_{\mathcal{S}_0} \vec{n}\cdot\vec{\hat{\mathcal{T}}} \cdot \boldsymbol{u}_{\mathcal{S}_M} \diff S.
\end{align}
The above expression is generic and can be evaluated for any microswimmer, provided the flow fields of the microswimmer near a planar wall and the equivalent auxiliary problem are known.

\paragraph{Far-field description.--}We employ our theoretical predictions to investigate the dynamics of an axisymmetric microswimmer, whose far-field flow is described as a combination of a force- and torque-dipole of strengths $\alpha_{\rm FD}$ and $\alpha_{RD}$, respectively~\cite{Spagnolie:2012}. Its instantaneous position $\vec{r}_0(t)=(x_{0}(t),y_{0}(t),h(t))^T$ and orientation $\vec{e}(t)=(\cos\vartheta(t)\cos \varphi(t),\cos\vartheta(t)\sin \varphi(t),\sin\vartheta(t))^T$, parametrized in terms of the pitch angle $\vartheta$ (with $\theta=0$ being parallel to $\mathcal{S}_0$) and polar angle~$\varphi$, obey the equations of motion: $\diff{\vec{r}}_0/\diff t =\vec{U}^{(0)}+\epsilon\vec{U}^{(1)}$ and
$\diff{\vec{e}}/\diff t=(\vec{\Omega}^{(0)}+\epsilon\vec{\Omega}^{(1)})\times \vec{e}$,
where the zeroth-order contributions, $\vec{U}^{(0)}$ and $\vec{\Omega}^{(0)}$, represent the translational and angular velocities near a planar wall~\cite{Spagnolie:2012, supp} and $\vec{U}^{(1)}$ and $\vec{\Omega}^{(1)}$ [Eq.~\eqref{eq:reciprocal}] are the contributions due to the deformation. We use parameters for a typical \textit{E.~Coli} bacterium, 
swimming at a velocity $U_{\rm free}\approx 22\, \mu {\rm m} {\rm s}^{-1}$ and having a body radius $a\approx\, 1\mu {\rm m}$. Its 
force- and torque-dipole strengths have been measured experimentally~\cite{Drescher:2011} and through simulations~\cite{Hu:2015}: ${\alpha}_{\rm FD}\approx\, 32\, {\rm \mu m^{3}\,s^{-1}}$ and  ${\alpha}_{\rm RD}\approx\, 25\, {\rm \mu m^{4}\,s^{-1}}$. Furthermore, we assume the deformation to be instantaneous compared to the swimming time scale, $\tau = a/U_{\rm free}$, so that $\partial_t\delta\approx 0$. The near-field and steric interactions of the microswimmer with the boundary are mimicked by a short-ranged repulsive force, so that the closest particle-surface distance is $h^{\star}=1.5a$~\cite{Spagnolie:2012,Poddar:2020,supp}. We study the trajectories of the microswimmers starting from a position $\vec{r}_0(0)=(0,0,4a)$ with pitch angles $\vartheta(0)\in [-\pi/2,\pi/2]$ and $\varphi(0)=0$ until the swimmer attains a stable orientation in the $xz$-plane, i.e. $\diff\vartheta/\diff t=-\Omega_{y}=0$ at $\vartheta=\vartheta^{\star}$. We further vary the elasto-viscous and FvK numbers, $\epsilon \in [0.01,0.1]$ and $\Gamma \in [0.01,10]$, respectively.

\begin{figure}[tp]
\centering
{\includegraphics[width = \columnwidth]{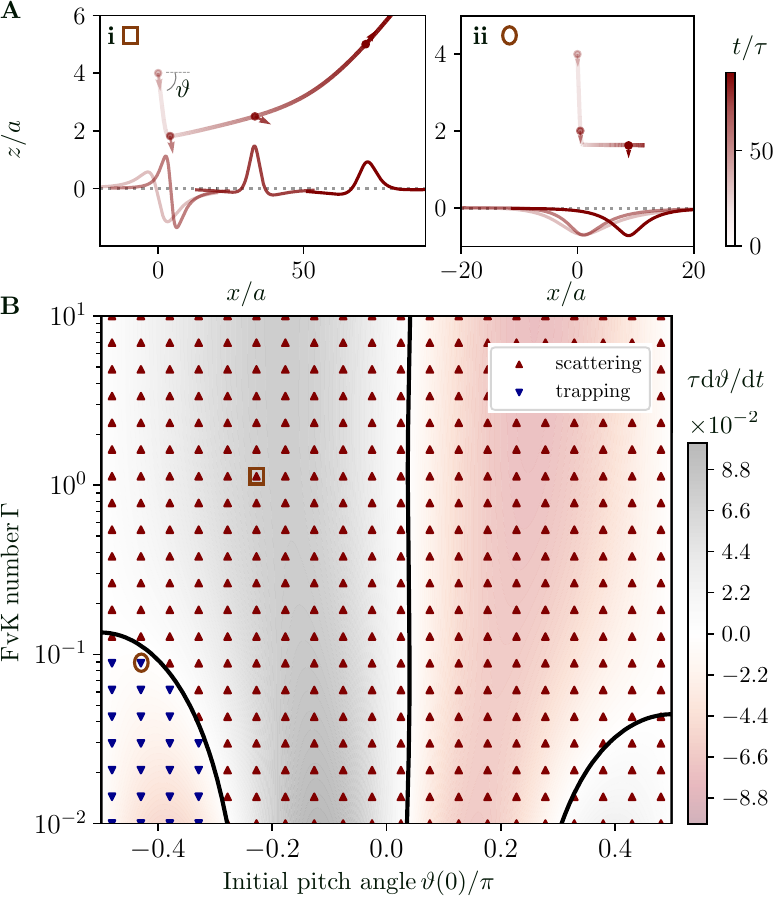}}
\caption{({\bf A}) Chronophotographies of a pusher-type microswimmer moving near a deformable boundary in the $xz-$plane. ({\bf A\,i})~Hydroelastic scattering event with parameters $\vartheta(0)=-0.23\pi$ and $\Gamma = 1.13$. ({\bf A\,ii})  Hydroelastic trapping event with parameters $\vartheta(0)=-0.43\pi$ and $\Gamma = 0.088$. The trajectories are color-coded by time, where $\tau=a/U_{\rm free}$ denotes the characteristic swimming time scale. The swimmer's orientation and the boundary deformation are shown at selected times. The latter are magnified by a factor of $10$ for visualization in ({\bf A\,i}). ({\bf B}) Phase diagram for a pusher as a function of the initial orientation~$\vartheta(0)$ and FvK number~$\Gamma$. Background colors indicate $\tau\diff\vartheta/\diff t=-\Omega_{y}(\epsilon,\Gamma,\vartheta(0),h^\star)$ and black lines correspond to the associated fixed points\,($\Omega_{y}=0$). The trajectories in ({\bf A}) correspond to the square and circle in the phase diagram. We use $\epsilon = 0.1$ in ({\bf A-B}).}\label{fig:pusher}
\end{figure}

\paragraph{Hydroelastic scattering of pushers.--} We begin our analysis by considering {\it E. coli}, i.e. a pusher-type microswimmer, oriented towards a deformable boundary at $\vartheta(0)=-0.23\pi$ with $\epsilon = 0.1$ and $\Gamma=1.13$. Importantly, we observe that the swimmer reorients and is scattered away from the boundary [Fig.~\ref{fig:pusher}{\bf A}(\textbf{i})]. This is primarily due to the deformation-induced velocity field giving rise to an angular velocity\,($\Omega_{y}$) rotating the swimmer. The shape of the deformation changes with the swimmer position and orientation: as the swimmer moves towards the surface, it pushes fluid towards it, leading to a valley, while it pulls fluid inwards from below, creating a hill. For a swimmer aligned parallel to $\mathcal{S}_0$, the deformation becomes symmetric. As it moves away from the boundary, the deformation shape changes to a valley at the back and a hill in front.

Our observation of hydroelastic scattering stands in stark contrast to the behavior of pushers aligning parallel to planar walls~\cite{Spagnolie:2012} -- a phenomenon that promotes surface trapping of bacteria~\cite{Berke:2008}. It is worth pointing out that due to the scattering the nature of the circular motion observed near planar walls~\cite{Lauga:2006} becomes negligible. This implies that the torque-dipole actuated angular velocity normal to the wall\,($\Omega_{z}$) is not decisive in the scattering trajectory for these parameters. Hence, it suffices to look at the trajectory of the swimmer in the $xz$-plane as in Fig.~\ref{fig:pusher}{\bf A}.

Moreover, it is paramount to note that the hydroelastic scattering appears reminiscent to the lift observed for their driven spherical counterparts~\cite{Rallabandi:2018}, yet the mechanism is profoundly different as the latter do not have a characteristic orientation. While the contribution to the deformation-induced velocity of spheres translating parallel to the elastic boundary becomes positive $U^{(1)}_z>0$, leading to the lift, the one of pushers is purely attractive $U^{(1)}_z<0$ for $\theta=0$ [Fig.~2\textbf{A} in SI~\cite{supp}], indicating  that the overall scattering of pushers away from the wall results from a hydroelastic reorientation. Further differences can be observed in the deformation which are symmetric for pushers aligned parallel to the wall [Fig.~\ref{fig:pusher}{\bf A}(\textbf{i})], but display asymmetric shapes for sedimenting spheres irrespective of their distance to the boundary~\cite{Rallabandi:2018}.

\paragraph{Hydroelastic trapping of pushers.--} Studying the trajectories of a pusher initially oriented towards the boundary at $\vartheta(0)=-0.43\pi$ and a smaller FvK number $\Gamma=0.088$, we find a notably different behavior: the microswimmer reorients towards the surface due to the hydroelastic interaction and finally ends up in a trapping state, with $\vartheta^{\star}=-\pi/2$. This can be rationalized by investigating Eq.~\eqref{helfrich} and the deformation shapes in [Fig.~\ref{fig:pusher}\textbf{A}(\textbf{ii})]. For $\Gamma \ll 1$, the bending resistance of the membrane dominates over the surface tension, with an increased penalty for higher-order gradients, producing deformations with smaller curvatures compared to situations with $\Gamma\gtrsim 1$. This appears to rotate the swimmer towards the boundary, leading to a symmetric deformation reminiscent of a pusher digging its own trap.

\paragraph{Phase diagram of pushers.--}To further characterize the emergence of these behaviors as a function of initial angles $\theta(0)$ and FvK numbers $\Gamma$ (keeping $\epsilon=0.1$ fixed), we plot a phase diagram where different phases are quantified based on their stationary orientation~$\theta^\star$\, [Fig.~\ref{fig:pusher}{\bf B}]. Our findings demonstrate that the hydroelastic scattering appears across various~$\vartheta(0)$ and~$\Gamma$, showing that it is a persistent feature for interactions between pushers and deformable boundaries. The phase diagram further reveals an extended trapping phase at small $\Gamma\lesssim 0.1$ and initial pitch angles close to $\theta(0)\sim -\pi/2$. To quantify the transition between both phases, we assume that the orientational dynamics close to the surface are important and thus we consider $\diff\vartheta/\diff t$ evaluated close to the surface at $h^{\star}$, indicated in the background of the phase diagram. Our results show that different stationary orientations emerge depending on $\Gamma$: for $\Gamma \gtrsim 0.1$ the only stable fixed point is $\theta^\star\sim 0.04\pi$, while for decreasing $\Gamma \lesssim 0.1$ the unstable fixed point at $\theta^\star=-\pi/2$ becomes stable (corresponding to the trapping state) and, akin a pitchfork bifurcation, two new unstable fixed points arise close to $\pm \pi/2$.  The unstable fixed point near $-\pi/2$  captures exactly the crossover between hydroelastic scattering and trapping in our phase diagram, emphasizing that the hydroelastic reorientation determines the overall dynamics.     

Finally, we note that the phase diagram remains independent of $h(0)$ [Fig.~5 in SI~\cite{supp}]. Also, for smaller elasto-viscous numbers, $\epsilon = 0.01$, the hydroelastic scattering and trapping phases disappear and we recover the in-plane circular motion. Increasing the elasto-viscous number slightly, $\epsilon=0.05$, the regime of scattering appears, while the trapping occurs for $\Gamma\lesssim 0.01$  [Fig.~3 in SI~\cite{supp}]. A substantial deformation seems to be necessary to observe these intriguing dynamics.

\begin{figure}[tp]
\centering
{\includegraphics[width = \columnwidth]{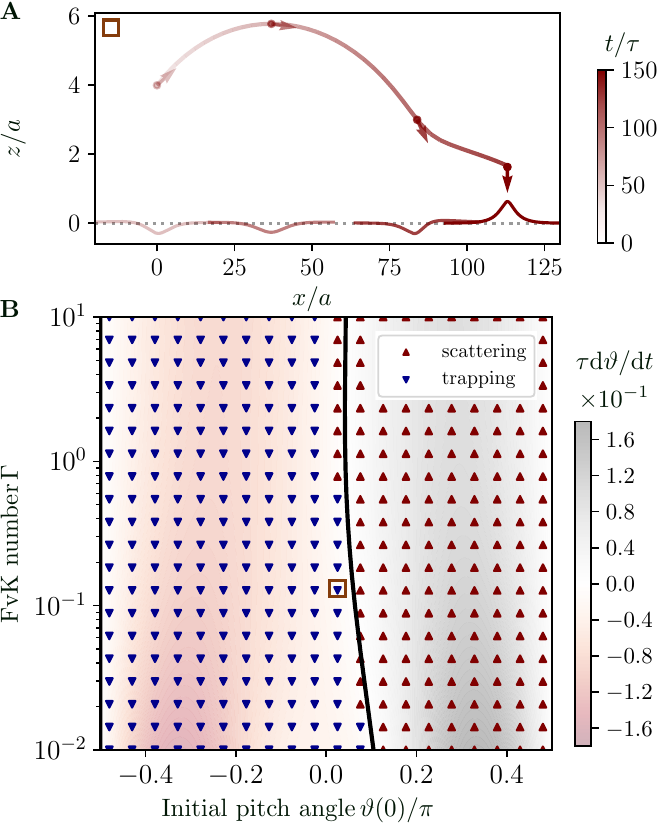}}
\caption{({\bf A})~Chronophotographies of a puller-type microswimmer moving near a deformable boundary: a hydroelastic trapping event. The parameters are $\vartheta(0)=0.02\pi$  and $\Gamma=0.13$.  The trajectories are color-coded by time, where $\tau=a/U_{\rm free}$ denotes the characteristic swimming time scale. The swimmer's orientation and the boundary deformation are shown at selected times. ({\bf B}) Phase diagram for a puller as a function of its initial orientation~$\vartheta(0)$ and FvK number~$\Gamma$. Background colors indicate $\diff\vartheta/\diff t$ at $h^{\star}$ and the black line corresponds to the associated unstable fixed points\,($\diff\vartheta/\diff t=0$). The trajectory in ({\bf A}) corresponds to the square in the phase diagram.  We use $\epsilon = 0.1$ in ({\bf A-B}). }\label{fig:puller}
\end{figure}

\paragraph{Enhanced trapping of pullers.--} While we have established a physical understanding of pusher-type microswimmers near a deformable boundary, we further investigate pullers with $\alpha_{\rm FD}\rightarrow -\alpha_{\rm FD}$ (and keeping $\alpha_{\rm RD}$) in terms of a phase plot for varying~$\vartheta(0)$ and~$\Gamma$. Most prominently, our results show that pullers, which are slightly oriented away from the boundary, exhibit enhanced overall trapping near the deformable boundary [Fig.~\ref{fig:puller}], in contrast to their scattering from a planar wall. A typical trajectory is depicted in Fig.~\ref{fig:puller}{\bf A}, indicating that the pullers first swim away from the boundary, before reorienting and moving towards it. In the trapping state, the swimmer pulls the boundary up creating a hill as opposed to the deformation produced by pushers.

The transition between trapping\,($\vartheta^{\star}=-\pi/2$) to a scattering state can be captured again by the unstable fixed points of the orientational dynamics $\diff\vartheta/\diff t$ at $h^{\star}$, as indicated in Fig~\ref{fig:puller}{\bf B}. Thus, the hydroelastic reorientation is the key ingredient for this intricate behavior, similar to the pusher case.

\paragraph{Summary and conclusions.--} Our study reveals a profound impact of a deformable boundary on the dynamics of microswimmers, giving rise to behaviors that are distinct to motion near rigid walls. Using a perturbation theory, we find that surface deformations can generate reorientations, effectively scattering pusher-type microswimmers away from the boundary and can enhance the overall trapping of pullers. Additionally, we observe that pushers, in certain parameter regimes, become trapped near deformable boundaries. 

Our results provide an analytical understanding of experimental observations of the trapping of active colloids~\cite{Vutukuri:2020, Fessler:2024} and bacteria~\cite{Takatori:2020} at the boundary of vesicles. In qualitative agreement with the finding that longer protrusions emerge for `flaccid' vesicles with low surface tension~\cite{Vutukuri:2020}, the trapping state in our work occurs for small  Föppl-von K{\'a}rm{\'a}n numbers. Besides, it lays the foundation for elucidating the emergence of collective effects due to the hydroelastic coupling~\cite{Takatori:2020, Vutukuri:2020, Fessler:2024, Azadbakht:2024}, in which deformation due to one influences the behavior of all other particles. This would thus complement recent simulations on vesicle deformations due to `dry' active agents~\cite{Paoluzzi:2016, Li:2019, Wang:2019} and pave the way towards gaining fundamental insights into the role of hydrodynamics. 

In this work we have considered an infinitely thin elastic boundary with fluid on one side. A natural extension is to include another fluid or a viscoelastic material on the other side~\cite{Rallabandi:2024}. Insights come from a theoretical study on rigid slender objects near a deformable liquid-liquid interface~\cite{Nambiar:2022}, where enhanced attraction, similar to our work, has been observed at long times. Furthermore, motion of reciprocal squirmers has been found due to the interplay of the time scales of the surface deformation and squirming modes~\cite{Shaik:2017}. Yet, the stationary dynamics of microswimmers near these interfaces, even in the far-field regime, as a function of material properties of the boundary remain to be explored.

Additionally, fluctuations at different levels, ranging from diffusion~\cite{Fares:2024} to short-range interactions~\cite{Kantsler:2013} to tumbling dynamics~\cite{Molaei:2014, Junot:2022, Kurzthaler:2024}, represent important aspects that could be incorporated in our framework and are expected to be highly relevant for the interactions of microswimmers with deformable boundaries in microbiological settings. In the future it will be important to develop numerical tools to ultimately relax the assumption of small and instantaneous boundary deformations and resolve the full non-linear hydroelastic  coupling, which is expected to reveal more enriching dynamics of the microswimmers.

\paragraph{Acknowledgments.--}C.K. gratefully acknowledges discussions with Howard A. Stone and Evgeniy Boyko. A.V. acknowledges funding from the Alexander von Humboldt foundation. 

\bibliography{literature}

\end{document}

%%%%%%%%%%%%%%%%%%%%%%%%%%%%%%%%%%%%%%%%%%%%%%%%%%%

\title{Supplemental Material: Hydroelastic Scattering and Trapping of Microswimmers}
\author{Sagnik Garai}
\affiliation{Max Planck Institute for the Physics of Complex Systems, N\"othnitzer Stra{\ss}e 38,
01187 Dresden, Germany}
\author{Ursy Makanga}
\affiliation{Max Planck Institute for the Physics of Complex Systems, N\"othnitzer Stra{\ss}e 38,
01187 Dresden, Germany}
\author{Akhil Varma}
\affiliation{Max Planck Institute for the Physics of Complex Systems, N\"othnitzer Stra{\ss}e 38,
01187 Dresden, Germany}
\author{Christina Kurzthaler}
\email{ckurzthaler@pks.mpg.de}
\affiliation{Max Planck Institute for the Physics of Complex Systems, N\"othnitzer Stra{\ss}e 38,
01187 Dresden, Germany}
\affiliation{Center for Systems Biology Dresden, Pfotenhauerstraße 108, 01307 Dresden, Germany}
\affiliation{Cluster of Excellence, Physics of Life, TU Dresden, Arnoldstraße 18, 01062 Dresden, Germany}

\maketitle
 
\tableofcontents

\section{Problem set-up \label{sec:problem-setup}}	
\begin{figure}[htp]
\centering
{\includegraphics[width = 0.7\columnwidth]{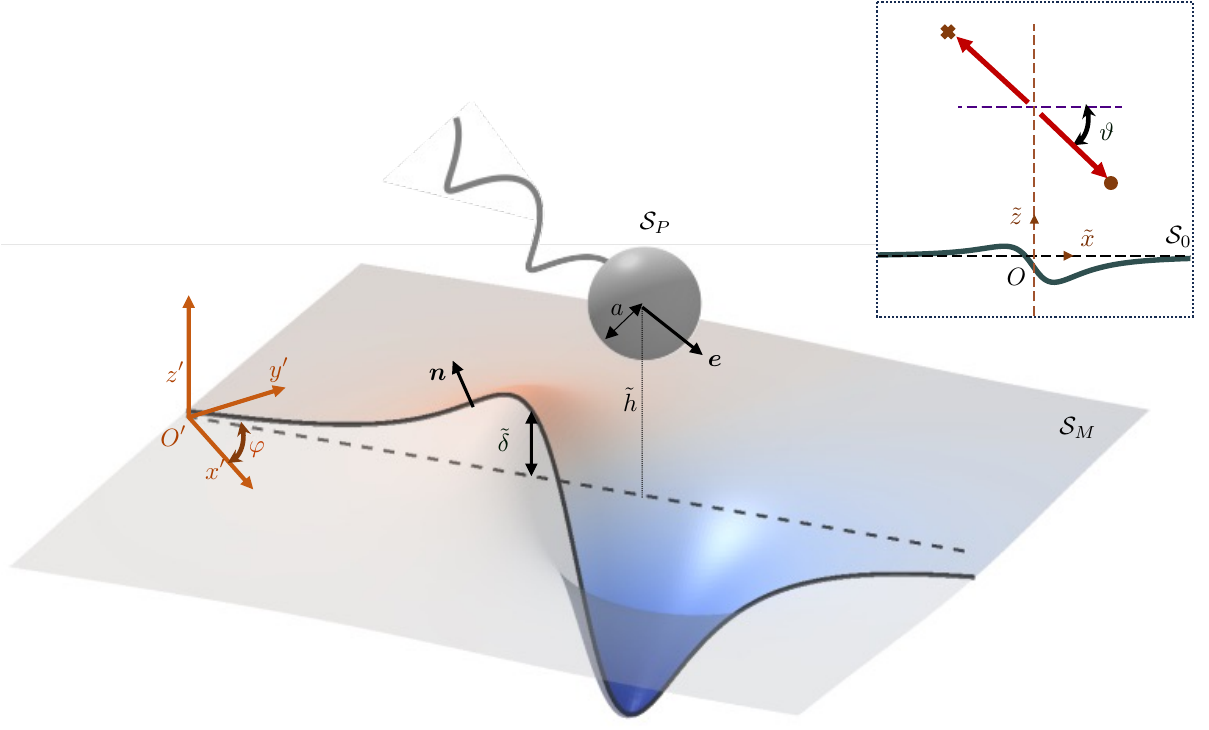}}
\caption{Model set-up of a swimming microorganism near a deformable boundary~($\mathcal{S}_{M}$). The deformation is denoted by $\tilde{\delta}$ with associated normal $\boldsymbol{n}$. A lab-fixed coordinate system is defined at $O'$\,(with coordinates\,$\{x',y',z'\}$). The undeformed surface $\mathcal{S}_{0}$ at $z'=0$ is considered as the reference surface (see also inset). The microswimmer, having a characteristic size $a$ and surface $\mathcal{S}_{P}$, propels along the direction $\boldsymbol{e}=(\cos\theta\cos\varphi,\cos\theta\sin\varphi,\sin\theta)$, where $\theta$ and $\varphi$ represent the angles with respect to the $z'-$ and $x'-$ axis, respectively. ({\it Inset})~Far-field model of the microswimmer, at height~$\tilde{h}$, represented as a combination of a force \,(red arrows) and torque dipole (indicated by the brown dot and cross). The center of the coordinate system $O$ (with coordinates\,$\{\tilde{x},\tilde{y},\tilde{z}\}$), which translates at a velocity $\Uparad$\,, is the projection of the microswimmer position to $\mathcal{S}_{0}$.}\label{fig1}
\end{figure}
Here, we present an extended version of our model description. For convenience, $\tilde{(\cdot)}$ represent dimensional quantities, while $(\cdot)$ denote their non-dimensional counterparts. 

We consider a microswimmer in an incompressible fluid of viscosity $\mu$ near an elastic boundary ($\mathcal{S}_M$), see Fig.~\ref{fig1}. The microswimmer creates a disturbance flow described by the spatially-varying fluid velocity field  $\boldsymbol{\tilde{u}}(\boldsymbol{\tilde{r}})$ and pressure field $\tilde{p}(\boldsymbol{\tilde{r}})$. These are governed in the low-Reynolds-number limit by the quasi-steady Stokes equations~\cite{Kim:2005, Happel:1983, Leal:2007}
\begin{align}\label{stokes}
\boldsymbol{\tilde{\nabla}} \tilde{p}=\mu\tilde{\nabla}^{2}\boldsymbol{\tilde{u}}\quad {\rm and}\quad \boldsymbol{\tilde{\nabla}}\cdot \boldsymbol{\tilde{u}}=0.
\end{align}
The swimming mechanism of the active agent is modeled in terms of a surface slip velocity $\boldsymbol{\tilde{u}}_{S}$, which describes the disturbance flows generated by, for example, non-reciprocal surface deformations of microorganisms~\cite{Lighthill:1952, Lauga:2020} or phoretic propulsion mechanisms~\cite{Golestanian:2005, Popescu:2016, Golestanian:2019, Michelin:2023}. As response to the surface distortions, the microswimmer translates at velocity $\boldsymbol{\tilde{U}}$ and rotates at an angular velocity $\boldsymbol{\tilde{\Omega}}$. For convenience, we decompose the microswimmer's velocity in terms of its velocity parallel and perpendicular to the planar reference surface $\mathcal{S}_0$ at $z'=0$, $\boldsymbol{\tilde{U}}=\boldsymbol{\tilde{U}}_\parallel+\boldsymbol{\tilde{U}}_\perp$. We further move to a reference frame $(O, \tilde{\boldsymbol{x}}, \tilde{\boldsymbol{y}}, \tilde{\boldsymbol{z}})$ on the undeformed reference surface $\mathcal{S}_{0}$, which translates with respect to the lab frame $(O', \boldsymbol{x}', \boldsymbol{y}', \boldsymbol{z}')$ at a velocity $\boldsymbol{\tilde{U}}_{\parallel}$ [Fig.~\ref{fig1}({\it inset})]. In this coordinate system, the swimmer is at a distance $\tilde{h}$ above the origin and the projection of the swimming direction $\vec{e}$ on the planar surface is along the $\tilde{x}-$axis. Then, the boundary condition (BC) on the surface of the swimmer reads
\begin{align}\label{bc1}
\boldsymbol{\tilde{u}}=\boldsymbol{\tilde{u}}_{S}+\Uperpd+\boldsymbol{\tilde{\Omega}}\times \boldsymbol{\tilde{R}} \quad {\rm on}\quad{\mathcal{S}_{P}}, 
\end{align}
where we abbreviated $\boldsymbol{\tilde{R}}= \boldsymbol{\tilde{r}}- (0,0,\tilde{h})$ and denote by $\boldsymbol{\tilde{r}}$  the vector pointing from $O$ to the surface of the microswimmer.  The surface of the deformable boundary is described by the  locus of all points $\boldsymbol{\tilde{r}}_{s}$ on the surface defined by $\tilde{F}(\boldsymbol{\tilde{r}}_{s},t)=0$\,\cite{Leal:2007}. A convenient description of the deformable boundary is the Monge gauge\,\cite{Deserno:2015}: $\tilde{F}(\boldsymbol{\tilde{r}}_{s},t)\equiv\tilde{z}-\tilde{\delta}(\boldsymbol{\tilde{r}}_\parallel,t)=0$, where $\tilde{\delta}(\boldsymbol{\tilde{r}}_\parallel,t)$ denotes the deformation and $\boldsymbol{\tilde{r}}_\parallel=(\tilde{x},\tilde{y})$ are the in-plane coordinates. The deformation of the boundary is governed by the stress balance~\cite{Helfrich:1973, Rallabandi:2018}
\begin{align}
\left(\kappa\tilde{\nabla}_{\parallel}^{4}-\Sigma\tilde{\nabla}_{\parallel}^{2}\right)\tilde{\delta}=\boldsymbol{\tilde{\sigma}}:\boldsymbol{n}\boldsymbol{n}\quad {\rm on}\quad{\mathcal{S}_{M}},\label{helfrich}
\end{align}
where $\tilde{\vec{\nabla}}_{\parallel}=\left(\partial_{\tilde{x}},\partial_{\tilde{y}}\right)$ is the in-plane gradient and $\boldsymbol{n}= \tilde{\grad} \tilde{F}/|\tilde{\grad} \tilde{F}|$ is the unit normal vector, pointing from the boundary into the fluid. The deformation and its normal derivative vanish at infinity. Further, $\kappa$ and $\Sigma$ denote the bending rigidity and the surface tension, respectively. 
The associated kinematic BC on $\mathcal{S}_M$ reads~\cite{Leal:2007}
\begin{align}
\frac{D\tilde{F}}{Dt}=\frac{\partial \tilde{F}}{\partial t}+\boldsymbol{\tilde{u}}\cdot \tilde{\grad} \tilde{F}=0\quad {\rm on}\quad{\mathcal{S}_{M}},
\label{bc3}
\end{align}
and a no-slip BC is prescribed on the boundary via
\begin{align}
(\boldsymbol{\tilde{u}}+\Uparad)
\cdot(\mathbb{I}-\boldsymbol{n}\boldsymbol{n})=\boldsymbol{0} \quad {\rm on}\quad{\mathcal{S}_{M}}, 
\end{align}
where $\mathbb{I}$ denotes the identity matrix. Far from the agent, the flow velocity vanishes. The latter implies $\tilde{\ubold}=-\Uparad$ on the bounding surface at infinity, $\mathcal{S}_\infty$. We further note that the neutrally-buoyant microswimmer is both  force- and torque-free,
\begin{align} 
\boldsymbol{\tilde{{F}}}_{H}=\int_{\mathcal{S}_{P}}\boldsymbol{{n}}\cdot\boldsymbol{\tilde{\sigma}}\,{\rm d}S=\boldsymbol{0}\,\, \quad {\rm and}\,\,\quad \boldsymbol{\tilde{{L}}}_{H}=\int_{\mathcal{S}_{P}}\tilde{\R} \times\left(\boldsymbol{{n}}\cdot\boldsymbol{\tilde{\sigma}}\right)\,{\rm d}S=\boldsymbol{0}, \label{ftfree}
\end{align}
where  $\boldsymbol{\tilde{\sigma}}=-\tilde{p}\mathbb{I}+\mu(\boldsymbol{\tilde{\nabla} \tilde{u}}+\boldsymbol{\tilde{\nabla} \tilde{u}}^{T})$ denotes the associated stress tensor and $\boldsymbol{{n}}$ is the unit normal pointing from the swimmer surface $\mathcal{S}_P$ into the fluid. 
		
It is convenient to non-dimensionalize the governing equations using the swimming velocity of the microswimmer in a free space $U_{\rm free}$, a typical length scale of the microswimmer $a$, e.g., the radius of its spherical head, and  $\mu U_{\rm free}/a$ as characteristic stress scale:
\begin{align}
\boldsymbol{\tilde{r}}=a\boldsymbol{r}, \quad \boldsymbol{\tilde{u}}=U_{\rm free}\boldsymbol{u}, \quad \tilde{p}=\frac{\mu \Uswim}{a}p, \quad \boldsymbol{\tilde{U}}=\Uswim\boldsymbol{U}, \quad \boldsymbol{\tilde{\Omega}}=\frac{\Uswim}{a}\boldsymbol{\Omega},\quad \tilde{F}=aF. 
\end{align}
We arrive at the non-dimensional set of equations 
\begin{subequations}
\begin{align}
{\nabla}^{2}\boldsymbol{u}=\boldsymbol{{\nabla}}p\quad {\rm and}\quad\boldsymbol{{\nabla}}\cdot \boldsymbol{u}=0,\label{ndstokes} \\
\left[ \frac{1}{1+\Gamma}\nabla_{\parallel}^{4}-\frac{\Gamma}{1+\Gamma}\nabla_{\parallel}^{2}\right]\delta=\epsilon\boldsymbol{\sigma}:\boldsymbol{n}\boldsymbol{n}\quad {\rm on}\quad{\mathcal{S}_{M}},\label{ndhelfrich}
\end{align}
\end{subequations}	
with associated BCs
\begin{subequations}
\begin{align}			
&\boldsymbol{u}=\boldsymbol{u}_{S}+\Uperpnd+\boldsymbol{\Omega}\times\R\quad {\rm on}\quad{\mathcal{S}_{P}},
&\qquad &\boldsymbol{u}=-\Uparand\quad {\rm on}\quad{\mathcal{S}_{\infty}},\label{ndbc2}\\
&\frac{1}{|\grad F|}\frac{\partial F}{\partial t}+\boldsymbol{u}\cdot \boldsymbol{n}=0\quad  \quad {\rm on}\quad{\mathcal{S}_{M}}, &\qquad &(\boldsymbol{u}+\Uparand)\cdot (\mathbb{I}-\boldsymbol{n}\boldsymbol{n})= \boldsymbol{0}\quad {\rm on}\quad{\mathcal{S}_{M}},\label{ndbc4}
\end{align}
\end{subequations}
and non-dimensional stress tensor  $\boldsymbol{\sigma}=-p\mathbb{I}+(\boldsymbol{{\nabla} u}+\boldsymbol{{\nabla} u}^{T})$, which serves as input for the non-dimensional hydrodynamic force, $\boldsymbol{{F}}_H$, and torque, $\boldsymbol{{L}}_H$ [Eq.~\eqref{ftfree}]. The non-dimensional numbers of our problem represent the generalized elasto-viscous number~\cite{Lee_Chadwick_Leal_1979} and the Föppl-von K{\'a}rm{\'a}n number:
\begin{align}
\epsilon= \frac{\mu a^{2}\Uswim}{\kappa+\Sigma a^{2}}\qquad {\rm and}\qquad \Gamma = \frac{\Sigma a^2}{\kappa}. \label{parameters}
\end{align}
In our work we vary the boundary properties ($\kappa$ and $\Sigma$) while keeping the swim speed~$U_{\rm free}$, the microswimmer length scale~$a$, and fluid viscosity~$\mu$ constant. Thus, for fixed $\epsilon$ we ensure that the sum $\kappa+\Sigma a^2$ remains constant.

 \section{Small-deformation limit: Perturbation approach \label{sec:perturbation}}
Assuming that the elastic resistance of the boundary is large compared to the viscous stress induced by the microswimmer, corresponding to small elasto-viscous numbers $\epsilon\ll1$, allows us considering only small deformations of the boundary and expanding the variables as
\begin{align}
\{\delta, \boldsymbol{u},p,\boldsymbol{\sigma},\boldsymbol{U}, \boldsymbol{\Omega}\} =  \{\delta^{(0)}, \boldsymbol{u}^{(0)},p^{(0)},\boldsymbol{\sigma}^{(0)}, \boldsymbol{U}^{(0)}, \boldsymbol{\Omega}^{(0)}\}+\epsilon\{\delta^{(1)},\boldsymbol{u}^{(1)},p^{(1)},\boldsymbol{\sigma}^{(1)},\boldsymbol{U}^{(1)}, \boldsymbol{\Omega}^{(1)}\}+\mathcal{O}(\epsilon^{2}),\label{expansion}
\end{align}
where $\delta^{(0)}=0$. Due to the linearity of the Stokes equations, we immediately find ${\nabla}^{2}\boldsymbol{u}^{(0)}=\boldsymbol{{\nabla}}p^{(0)}$ and $\boldsymbol{{\nabla}}\cdot \boldsymbol{u}^{(0)}=0$ and ${\nabla}^{2}\boldsymbol{u}^{(1)}=\boldsymbol{{\nabla}}p^{(1)}$ and $\boldsymbol{{\nabla}}\cdot \boldsymbol{u}^{(1)}=0$, respectively. In what follows, we derive the perturbation expansions of the stress balance on $\mathcal{S}_M$ and the associated BCs. 

Therefore, we first note that the small-deformation expansion allows us to expand the velocity on $\mathcal{S}_{M}$ in terms of a Taylor series about the reference surface $\mathcal{S}_0$ at $z=0$~\cite{Leal:2007, Kamrin:2010, Rallabandi:2018}, 
\begin{align}
\boldsymbol{u}=\boldsymbol{u}(z=\delta)=\boldsymbol{u}(z=0)+\left.\delta\frac{\partial \boldsymbol{u}}{\partial z}\right|_{z=0}+\mathcal{O}(\delta^2). 
\label{ndbcexp1}
\end{align}
Inserting Eq.~\eqref{expansion} into Eq.~\eqref{ndbcexp1} translates to 
\begin{align}
\boldsymbol{u}^{(0)}+\epsilon\boldsymbol{u}^{(1)}+\mathcal{O}(\epsilon^2)=\boldsymbol{u}^{(0)}(z=0)+\epsilon \boldsymbol{u}^{(1)}(z=0)+\epsilon \delta^{(1)}\frac{\partial \boldsymbol{u}^{(0)}}{\partial z}\Bigr|_{z=0}+\mathcal{O}(\epsilon^2), %\quad &{\rm on}\quad{\mathcal{S}_{M}}. 
\label{ndbcexp2}
\end{align}
This procedure permits transforming the problem of solving the Stokes  equations  with the BCs coupled to the configuration of the surface $\mathcal{S}_{M}$\,($z=\delta(\boldsymbol{r}_\parallel)$) into equivalent problems on the undeformed surface $\mathcal{S}_0$. 

\subsection{Perturbation expansion of the boundary conditions\label{app:BC}}
The BC on the deformable surface [Eq.~\eqref{ndbc4}] is perturbatively expanded. Substituting the Taylor series of the velocity on the boundary $\mathcal{S}_{M}$ about the undeformed surface $\mathcal{S}_{0}$\,[Eq.~\eqref{ndbcexp2}] into the kinematic and the no-slip BCs [Eq.~\eqref{ndbc4}], we obtain
\begin{subequations}
    \begin{align}\label{kinematicbc}
      - \epsilon\frac{ \partial{\delta}^{(1)}}{\partial t}+\left(\boldsymbol{u}^{(0)}(z=0)+\epsilon\boldsymbol{u}^{(1)}(z=0)+\left.\epsilon \frac{\partial \boldsymbol{u}^{(0)}}{\partial z}\right|_{z=0}\right)\cdot\left(\z-\epsilon\grad \delta^{(1)}\right)+\mathcal{O}(\epsilon^{2})=0, 
    \end{align}
and 
    \begin{align}\label{noslipbc}
        \left(\boldsymbol{u}^{(0)}(z=0)+\epsilon\boldsymbol{u}^{(1)}(z=0)+\left.\epsilon \delta^{(1)}\frac{\partial\boldsymbol{u}^{(0)}}{\partial z} \right|_{z=0}+\Uparand^{(0)}+\epsilon\Uparand^{(1)}\right)\cdot \left(\mathbb{I}-\z\z+\epsilon\z\grad\delta^{(1)}+\epsilon\grad\delta^{(1)}\z\right)+\mathcal{O}(\epsilon^{2})=\boldsymbol{0}, 
    \end{align}
\end{subequations}
on $\mathcal{S}_0$, respectively.
Collecting the terms in zeroth-order of $\epsilon$, we have for the kinematic condition [Eq.~\eqref{kinematicbc}]
\begin{subequations}
    \begin{align}
        \boldsymbol{u}^{(0)}(z=0)\cdot \z=0 \quad {\rm on}\quad \mathcal{S}_{0} \label{eq:kinematic_0}
    \end{align}
    and for the no-slip condition\,[Eq.~\eqref{noslipbc}]
    \begin{align}
     \left(\boldsymbol{u}^{(0)}(z=0)+\Uparand^{(0)}\right)\cdot (\mathbb{I}-\z\z)=\boldsymbol{0}\quad {\rm on}\quad \mathcal{S}_{0},
    \end{align}
\end{subequations}
leading to the zeroth-order BC: $\vec{u}^{(0)}=-\vec{U}_\parallel^{(0)}$ on $\mathcal{S}_{0}$. 
For the first-order in $\epsilon$-terms, we have for the kinematic BC\,[Eq.~\eqref{kinematicbc}]
\begin{subequations}
    \begin{align}\label{firstkinematicbc}
        -\frac{ \partial{\delta}^{(1)}}{\partial t}+\left(\boldsymbol{u}^{(1)}(z=0)+\delta^{(1)}\frac{\partial \boldsymbol{u}^{(0)}}{\partial z}\right)\cdot \z+\vec{U}_\parallel^{(0)}\cdot \grad \delta^{(1)}=0 \quad {\rm on}\quad \mathcal{S}_{0}
    \end{align}
    and for the no-slip condition\,[Eq.~\eqref{noslipbc}]
    \begin{align}\label{firstnoslipbc}
     \left(\boldsymbol{u}^{(1)}(z=0)+\delta^{(1)}\frac{\partial\boldsymbol{u}^{(0)}}{\partial z} +\Uparand^{(1)}\right)\cdot (\mathbb{I}-\z\z)=\boldsymbol{0} \quad {\rm on}\quad \mathcal{S}_{0}.
    \end{align}
\end{subequations}
By definition $\Uparand^{(1)}\cdot \z=0$, which we substitute into Eq.~\eqref{firstnoslipbc}. From the zeroth-order BC [Eq.~\eqref{eq:kinematic_0}] and the continuity equation ($\grad\cdot\boldsymbol{u}^{(0)}=0$), we have $\partial_z u^{(0)}_{z} |_{z=0}$=0, which we substitute into Eqs.~\eqref{firstkinematicbc} and~\eqref{firstnoslipbc}. Thus, we obtain
\begin{align}
   \boldsymbol{u}^{(1)}(z=0)+\delta^{(1)}\frac{\partial\boldsymbol{u}^{(0)}}{\partial z} +\Uparand^{(1)}-(\boldsymbol{u}^{(1)}(z=0) \cdot \z )\z=\boldsymbol{0} \quad {\rm on}\quad \mathcal{S}_{0}.
\end{align}
Substituting for $\boldsymbol{u}^{(1)}(z=0)\cdot \z$ from Eq.~\eqref{firstkinematicbc}, yields 
\begin{align}
  \boldsymbol{u}^{(1)}(z=0)+\delta^{(1)}\frac{\partial\boldsymbol{u}^{(0)}}{\partial z} +\Uparand^{(1)}+\left( -\frac{ \partial{\delta}^{(1)}}{\partial t}+\boldsymbol{U}^{(0)}_\parallel\cdot \grad\delta^{(1)}\right)\z=\boldsymbol{0} \quad {\rm on}\quad \mathcal{S}_{0}, \label{eq:BC_SM}
\end{align}
which gives Eq.~\eqref{ndbc31}, after simplification.

\subsection{Perturbation expansion of the stress balance at the boundary}
We insert the perturbation expansion [Eq.~\eqref{expansion}] into the equation for the boundary [Eq.~\eqref{ndhelfrich}]. Therefore, we first note that the expansion of the normal $\vec{n}$ yields
\begin{subequations}
	\begin{align}\label{normal}
		\boldsymbol{n}&=\frac{\grad F}{|\grad F|}
    =\left[1+\epsilon^{2}\{(\partial_{x}\delta^{(1)}(x,y))^{2}+(\partial_{y}\delta^{(1)}(x,y))^{2}\}+\mathcal{O}(\epsilon^{4})\right]^{-\frac{1}{2}}\left[\boldsymbol{\hat{z}}-\epsilon\boldsymbol{\nabla}_{\parallel}\delta^{(1)}(x,y)+\mathcal{O}(\epsilon^{2})\right],\\%\boldsymbol{\nabla}_{\parallel}\delta^{(1)}(x,y)\right\}, \\
		&= \boldsymbol{\hat{z}}-\epsilon\boldsymbol{\nabla}_{\parallel}\delta^{(1)}(x,y)+\mathcal{O}(\epsilon^{2}).%\quad{\rm where}\quad \boldsymbol{\nabla}_{\parallel}=\x \partial_{x} + \y \partial_{y}.
	\end{align}
 \end{subequations}
Expanding the stress tensor $\boldsymbol{\sigma}$ in a Taylor series about the undeformed surface $\mathcal{S}_{0}$
\begin{align}\label{stress_taylor}
    \boldsymbol{\sigma}(z=\epsilon\delta^{(1)})=\boldsymbol{\sigma}^{(0)}(z=0)+\epsilon \boldsymbol{\sigma}^{(1)}(z=0)+\epsilon\delta^{(1)}\frac{\partial \sigma^{(0)}}{\partial z}+\mathcal{O}(\epsilon^{2})\quad {\rm on}\quad \mathcal{S}_{0},
\end{align}
and substituting Eq.~\eqref{normal} and Eq.~\eqref{stress_taylor} into the right-hand side of Eq.~\eqref{ndhelfrich}, we find the leading-order expression
\begin{subequations}
	\begin{align}
		\epsilon \boldsymbol{\sigma}:\boldsymbol{n}\boldsymbol{n}&=\epsilon\left[-p^{(0)}\mathbb{I}+2\mathbf{E}^{(0)}+\mathcal{O}(\epsilon)\right]:\left[ \boldsymbol{\hat{z}}-\epsilon\boldsymbol{\nabla}_{\parallel}\delta^{(1)}(x,y)+\mathcal{O}(\epsilon^{2})\right]\left[\boldsymbol{\hat{z}}-\epsilon\boldsymbol{\nabla}_{\parallel}\delta^{(1)}(x,y)+\mathcal{O}(\epsilon^{2})\right]\quad {\rm on}\quad \mathcal{S}_{0}, \\
		&=\epsilon\left[-p^{(0)}+2\frac{\partial u^{(0)}_{z}}{\partial z}\right]_{z=0}+\mathcal{O}(\epsilon^{2}) \quad {\rm on}\quad \mathcal{S}_{0}.\label{stress_surface0}
	\end{align}
 \end{subequations}
 Using the continuity equation ($\boldsymbol{{\nabla}}\cdot \boldsymbol{u}^{(0)}=0$) and the no-slip condition ($\boldsymbol{u}^{(0)}=-\boldsymbol{U}^{(0)}_\parallel$  on $z=0$) it is evident that $\partial_z u^{(0)}_{z}|_{z=0}=0$. We thus arrive at 
	\begin{align}
		\epsilon \boldsymbol{\sigma}:\boldsymbol{n}\boldsymbol{n}|_{\mathcal{S}_{0}}=-\epsilon p^{(0)}|_{z=0}+\mathcal{O}(\epsilon^{2}).\label{stress_rigid}
	\end{align}

\subsection{Zeroth-order velocity field and leading-order deformation}
At leading order in $\epsilon$, the velocity and pressure fields, $\boldsymbol{u}^{(0)}$ and $p^{(0)}$, obey the  Stokes equations, $\nabla^{2}\boldsymbol{u}^{(0)}=\boldsymbol{\nabla}p^{(0)}$ and  $\boldsymbol{\nabla}\cdot \boldsymbol{u}^{(0)}=0$, subject to the BCs
\begin{align}	
\boldsymbol{u}^{(0)}=\boldsymbol{u}_{S}+\Uperpnd^{(0)}+\boldsymbol{\Omega}^{(0)}\times\R\quad {\rm on}\quad{\mathcal{S}_{P}},\qquad
\boldsymbol{u}^{(0)}=-\Uparand^{(0)}\,\,\quad {\rm on}\quad{\mathcal{S}_{\infty}},\quad \text{and} \quad
\boldsymbol{u}^{(0)}=-\Uparand^{(0)}\quad {\rm on}\quad{\mathcal{S}_{0}}.\label{ndbc30}
\end{align}
The zeroth-order problem represents the velocity field produced by a microswimmer near a planar wall\,($\mathcal{S}_0$), which has been solved analytically, for example, in the far-field regime in terms of singularity solutions~\cite{Spagnolie:2012} (see Sec.~\ref{sec:far-field}) and for the squirmer model using a bispherical coordinate representation~\cite{Poddar:2020}. 
  
The zeroth-order problem is coupled to the first-order problem through the deformation, which enters the BCs [Eq.~\eqref{eq:BC_SM}]. Thus, \,$p^{(0)}(\boldsymbol{r}_{\parallel},z=0)$ is substituted into Eq.~\eqref{stress_rigid} to solve for the first-order deformation~$\delta^{(1)}(\boldsymbol{r}_{\parallel})$.  It is convenient to move to Fourier space~\cite{Rallabandi:2018}, with $f(\boldsymbol{k})=\int f(\boldsymbol{r}_{\parallel})\exp(-i\boldsymbol{k}\cdot\boldsymbol{r}_{\parallel}) \diff^{2}\boldsymbol{r}_{\parallel}$ for $f=\delta^{(1)},p^{(0)}$, yielding  
\begin{align}
\delta^{(1)}(\boldsymbol{k})=- \left.\dfrac{1+\Gamma}{k^{4}+\Gamma k^{2}}p^{(0)}(\boldsymbol{k})\right|_{z=0},\label{defft1}
\end{align}
where we introduced the wavenumber $k=|\boldsymbol{k}|$. A backtransform then provides the leading-order deformation $\delta^{(1)}(\boldsymbol{r}_\parallel)$. 
		
\subsection{Reciprocal relation for the deformation-induced swimming velocities}
The zeroth-order velocity field $\vec{u}^{(0)}$ and the first-order deformation $\delta^{(1)}$ serve as input for deriving the leading-order contribution to the deformation-induced velocities, $\vec{U}^{(1)}$ and $\vec{\Omega}^{(1)}$, of the microswimmer.  Following our perturbation scheme, the first-order problem obeys the Stokes equations, $\nabla^{2}\boldsymbol{u}^{(1)}=\boldsymbol{\nabla}p^{(1)}$ and $\boldsymbol{\nabla}\cdot \boldsymbol{u}^{(1)}=0$,  with BCs
\begin{subequations}
\begin{align}	&\boldsymbol{u}^{(1)}=\Uperpnd^{(1)}+\boldsymbol{\Omega}^{(1)}\times\R, \quad{\rm on}\quad \mathcal{S}_P, \qquad 
\boldsymbol{u}^{(1)}=-\Uparand^{(1)}\quad {\rm on}\quad \mathcal{S}_\infty, \quad\text{and}\label{ndbc21}\\
&\boldsymbol{u}^{(1)}=-\Uparand^{(1)}+\vec{u}_{\mathcal{S}_M}\quad {\rm on}\quad {\mathcal{S}_0}\quad{\rm where}\quad \vec{u}_{\mathcal{S}_M}=\frac{\partial \delta^{(1)}}{\partial t}\z- \left.\delta^{(1)}\frac{\partial \boldsymbol{u}^{(0)}}{\partial z}\right|_{z=0} - \Uparand^{(0)}\cdot \grad_{\parallel}\delta^{(1)}\z.\label{ndbc31}
\end{align}
\end{subequations} 
We note that the solution to the first-order problem depends on the solution to the zeroth-order problem through the BC on~$\mathcal{S}_0$ [Eq.~\eqref{ndbc31}]. To compute the deformation-induced swimming velocities, $\boldsymbol{U}^{(1)}$ and $\boldsymbol{\Omega}^{(1)}$, we follow the approach of Ref.~\cite{Rallabandi:2018} and rely on the Lorentz reciprocal theorem~\cite{lorentz1897general, Happel:1983, Leal:2007, Masoud:2019}. We thereby circumvent computing the full velocity field $\boldsymbol{u}^{(1)}$. In particular, we introduce an auxiliary (known) problem with velocity field and stress tensor \,$\{\boldsymbol{\hat{u}},\boldsymbol{\hat{\sigma}}\}$ of a  passive object of the same shape as the microswimmer, moving near a planar (no-slip) wall with velocities $\boldsymbol{\hat{U}}$ and $\boldsymbol{\hat{\Omega}}$, under the application of an external force $\boldsymbol{\hat{F}}$ and torque $\boldsymbol{\hat{L}}$. The associated BCs are: $\boldsymbol{\hat{u}}=-\boldsymbol{\hat{U}}_\parallel$ on $\mathcal{S}_{\infty}$ and $\mathcal{S}_0$, and $\boldsymbol{\hat{u}}=\Uauxperp+\boldsymbol{\hat{\Omega}}\times\R$ on $\mathcal{S}_{P}$. The Lorentz reciprocal theorem then relates the first-order problem to the auxiliary problem via
\begin{align}
\int_{\mathcal{S}_{P}\cup \mathcal{S}_{0} \cup \mathcal{S}_{\infty}} \boldsymbol{n}\cdot \boldsymbol{\sigma}^{(1)}\cdot \boldsymbol{\hat{u}} \,\diff S=\int_{\mathcal{S}_{P}\cup \mathcal{S}_{0} \cup \mathcal{S}_{\infty}} \boldsymbol{n}\cdot \boldsymbol{\hat{\sigma}}\cdot \boldsymbol{u}^{(1)}\, \diff S. \label{reciprocal}
\end{align}
Inserting the BCs  of the main [Eqs.~\eqref{ndbc21}-\eqref{ndbc31}] and auxiliary problem into Eq.~\eqref{reciprocal}, leads to
\begin{align}
\begin{split}
&-\Uauxpara \cdot\int_{\mathcal{S}_{\infty}\cup \mathcal{S}_{0}} \boldsymbol{n}\cdot \sigmand^{(1)} \,\diff S+\Uauxperp\cdot\int_{\mathcal{S}_{P}} \boldsymbol{n}\cdot \sigmand^{(1)} \diff S + \boldsymbol{\hat{\Omega}} \cdot \int_{\mathcal{S}_{P}} \R\times\boldsymbol{n}\cdot \sigmand^{(1)} \,\diff S=\\ 
&-\Uparand^{(1)} \cdot\int_{\mathcal{S}_{\infty} \cup \mathcal{S}_0} \boldsymbol{n}\cdot \sigmahat \,\diff S + \Uperpnd^{(1)}\cdot\int_{\mathcal{S}_{P}} \boldsymbol{n}\cdot \sigmahat \,\diff S +\omegaboldnd^{(1)}\cdot\int_{\mathcal{S}_{P}} \R\times\boldsymbol{n}\cdot \sigmahat \,\diff S 
+\int_{\mathcal{S}_{0}} \boldsymbol{n}\cdot \sigmahat\cdot \vec{u}_{\mathcal{S}_M} \,\diff S \label{reciprocal2}   
\end{split}
\end{align}
Noting that for both the main and the auxiliary problem the total hydrodynamic force over the entire domain vanishes
\begin{align}
\int_{\mathcal{S}_{P}\cup \mathcal{S}_{0}\cup \mathcal{S}_{\infty} } \boldsymbol{n}\cdot \sigmahat \ \diff S=\boldsymbol{0}\quad {\rm and}\quad \int_{\mathcal{S}_{P}\cup \mathcal{S}_{0}\cup \mathcal{S}_{\infty} }  \boldsymbol{n}\cdot \sigmand^{(1)} \ \diff S=\boldsymbol{0},
\end{align}
and that the swimmer is force- and torque-free~[Eq.~\eqref{ftfree}], we find that Eq.~\eqref{reciprocal2} reduces to
\begin{align}\label{reciprocal3}
0=\boldsymbol{U}^{(1)}\cdot\int_{\mathcal{S}_{P}} \boldsymbol{n}\cdot \sigmahat \,\diff S&+\boldsymbol{\Omega}^{(1)}\cdot\int_{\mathcal{S}_{P}} \R\times\boldsymbol{n}\cdot \sigmahat \,\diff S +\int_{\mathcal{S}_{0}} \boldsymbol{n}\cdot \sigmahat\cdot \vec{u}_{\mathcal{S}_M}\,\diff S,    
\end{align}
which further simplifies to
\begin{align}\label{reciprocal4}
\boldsymbol{U}^{(1)}\cdot\boldsymbol{\faux}_{\rm H}+\boldsymbol{\Omega}^{(1)}\cdot\boldsymbol{\laux}_{\rm H}=-\int_{\mathcal{S}_{0}} \, \boldsymbol{n}\cdot \boldsymbol{\hat{\sigma}}\cdot\boldsymbol{u}_{\mathcal{S}_M} \diff S,
\end{align}
which, by using $\vec{\hat{F}}_H=-\vec{\hat{F}}$ and $\vec{\hat{L}}_H=-\vec{\hat{L}}$, results in Eq.~(3) of the main text.  

We note that Eqs.~(5)-(7) of  the main text can be used to formulate a general expression for the deformation induced velocities up to different orders of the perturbation expansion. In particular, for swimmers, whose self-propulsion mechanism is modeled in terms of a surface slip $\boldsymbol{u}_{\rm S}$, the generalized velocities $\boldsymbol{Q}= (\boldsymbol{U}, \boldsymbol{\Omega})^T$ can be expressed as
\begin{align}
    \boldsymbol{Q} &= \boldsymbol{\hat{\mathcal{R}}}^{-1} \cdot 
 \left[-\int_{\mathcal{S}_P}  \vec{n}\cdot\vec{\hat{\mathcal{T}}} \cdot \boldsymbol{u}_{S} \ \diff S+\epsilon \int_{\mathcal{S}_0}  \vec{n}\cdot\vec{\hat{\mathcal{T}}} \cdot \boldsymbol{u}_{\mathcal{S}_M} \ \diff S\right] + \mathcal{O}(\epsilon^2),
\end{align}
where  the definitions of $\vec{\hat{\mathcal{T}}}$ and $\boldsymbol{\hat{\mathcal{R}}}$ follow from the main text.
\section{Far-field hydrodynamics of a microswimmer near a deformable boundary \label{sec:far-field}}
In this section, we first provide details to the far-field model [Sec.~\ref{sec:far-field_wall}]. In Secs.~\ref{sec:deformation} and~\ref{sec:velocities} we provide details of the calculations of the deformation and deformation-induced velocities, respectively. 

\subsection{Far-field model of the microswimmer near a planar wall\label{sec:far-field_wall}}
The axisymmetric microswimmer is modelled as a combination of a force and torque dipole, where the forces and torques are directed along the dipole  $l\vec{e}/a$ of length $l$. These are obtained using the Green's function for a point force in a Stokes flow. For the sake of completion we outline here the solution to the flow and the pressure fields of the microswimmer near a planar, no-slip surface~\cite{Chwang:1975,Spagnolie:2012,Christina:2021}. First, we show that the free-space Green's function of the dipole terms are readily obtained from the Green's function of a point force (Stokeslet) solution of the Stokes equation. Subsequently, we trace the same formalism to derive the respective Green's function near a flat rigid surface.
Following previous work we use a lab based coordinate system at $O'$\,(with coordinates\,$\{x',y',z'\}$).

The Green's function to the Stokes equation in free space is obtained by placing a point force at $\boldsymbol{r}'_{0}=(0,0,h)$ with respect to some coordinate system. In non-dimensional form, this is given by\,:
\begin{align}\label{stokeslet}
\nabla^{2}\boldsymbol{u}'-\boldsymbol{\nabla}p'=\left(\frac{F}{\mu a\U}\right) \boldsymbol{e}\delta(\boldsymbol{r}'-\boldsymbol{r}'_{0}),\quad{\rm with}\quad \boldsymbol{u}'=\boldsymbol{0}\quad{\rm on}\quad\mathcal{S}_{\infty},\,\mathcal{S}_{0}, 
\end{align}
where $\boldsymbol{e}=\cos\theta\x'+\sin\theta\z'$ is a unit vector and $F$ is the magnitude of the point force. Redefining $\boldsymbol{R}'=\boldsymbol{r}'-\boldsymbol{r}_{0}$, the solution of the velocity and pressure fields, with $\vec{u}'\to\vec{0}$ far from the point force, are\,\cite{Happel:1983,Kim:2005} 
\begin{subequations}
\begin{align}
\boldsymbol{u}'&\equiv\boldsymbol{u}'_{\rm F}(\boldsymbol{e};\boldsymbol{R}')=\frac{F}{8\pi\mu a\U}\boldsymbol{e}\cdot\left[\frac{\boldsymbol{\mathbb{I}}}{R'}+\frac{\boldsymbol{R}'\boldsymbol{R}'}{R'^{3}}\right]
=\frac{F}{8\pi\mu a\U}\boldsymbol{G}_{\rm F}(\boldsymbol{R}';\boldsymbol{e})\\
p'&\equiv p'_{\rm F}(\boldsymbol{e};\boldsymbol{R}')=\frac{F}{8\pi\mu a\U} \vec{e}\cdot\frac{2\boldsymbol{R}'}{R'^{3}}
=\frac{F}{8\pi\mu a\U}P_{\rm F}(\boldsymbol{R}';\boldsymbol{e})
\end{align}
\end{subequations}
respectively, where $\boldsymbol{G}_{\rm F}$ and $P_{\rm F}$ are the Green's functions. The axisymmetric force-dipole solution is readily obtained from the point force by placing two point forces along $\e$ and $-\e$ respectively, separated by the non-dimensional dipole length $l/a$ along $\e$.  For the right-hand-side of Eq.~\eqref{stokeslet} we have
\begin{align}
\frac{F}{\mu\U a}\boldsymbol{e}\left[\delta\left(\boldsymbol{r}'-\left\{\boldsymbol{r}'_{0}+\frac{l}{2a}\boldsymbol{e}\right\}\right)-\delta\left(\boldsymbol{r}'-\left\{\boldsymbol{r}'_{0}-\frac{l}{2a}\boldsymbol{e}\right\}\right)\right]=-\frac{Fl}{\mu\U a^{2}}\boldsymbol{e}\left(\boldsymbol{e}\cdot \boldsymbol{\nabla}_{\boldsymbol{r}'-\boldsymbol{r}'_{0}}\right)\delta\left(\boldsymbol{r}'-\boldsymbol{r}'_{0}\right)+\mathcal{O}\left(\frac{l^{2}}{a^{2}}\right).
\end{align}
Thus, the corresponding solutions to the velocity and pressure fields, $\boldsymbol{u}'_{\rm FD}=Fl/(8\pi \mu \U a^{2})\boldsymbol{G}_{\rm FD}$ and $p'_{\rm FD}=Fl/(8\pi \mu \U a^{2})P_{\rm FD}$, respectively, are obtained from the directional derivative of the point-force Green's function: $\boldsymbol{G}_{\rm FD}(\boldsymbol{R}';\e,\e)=-(\e\cdot\grad_{\boldsymbol{R}'})\boldsymbol{G}_{\rm F}(\boldsymbol{R}';\e)$ and $P_{\rm FD}(\boldsymbol{R}';\e,\e)=-(\e\cdot\grad_{\boldsymbol{R}'})P_{\rm F}(\boldsymbol{R}';\e)$. 

Similarly, the solution for the point torque (rotlet) is obtained by solving Eq.~\eqref{stokeslet} with $L/(a^{2}\mu\U)\boldsymbol{\nabla}\times \boldsymbol{e}\delta(\boldsymbol{R}')$\,\cite{Chwang:1975,Spagnolie:2012} on the right-hand-side, where $L$ is the magnitude of the torque.  The corresponding Green's function is given by $\boldsymbol{G}_{\rm R}(\boldsymbol{R}';\e)=(1/2)\grad_{\boldsymbol{R}'}\times\boldsymbol{G}_{\rm F}(\boldsymbol{R}';\e)$ and $P_{\rm R}(\boldsymbol{R}';\e,\e)=0$. As in the case of a force dipole, placing two point torques about $\e$ and $-\e$ separated along $l\e/a$ on the right-hand-side of Eq.~\eqref{stokeslet} generates the Green's function for the torque dipole. Thus, the solution for the velocity and pressure fields for the torque dipole is given by  $\boldsymbol{u}'_{\rm RD}=Ll/(8\pi \mu \U a^{3})\boldsymbol{G}_{\rm RD}=-(Ll/8\pi \mu \U a^{3})(\e\cdot\grad_{\boldsymbol{R}'})\boldsymbol{G}_{\rm R}(\boldsymbol{R}';\e)$ and $p'_{\rm RD}=-Ll/(8\pi \mu \U a^{2})(\e\cdot\grad_{\boldsymbol{R}'})P_{\rm R}(\boldsymbol{R}';\e)$ respectively. The Green's function for the force and torque dipole are of the form
\begin{subequations}
    \begin{align}
        \boldsymbol{G}_{\rm FD}(\R';\e,\e)&=-\frac{\boldsymbol{R}'}{R'^{3}}+\frac{3(\boldsymbol{e}\cdot \boldsymbol{R}')^{2}\boldsymbol{R}'}{R'^{5}}, 
		& \quad
		& P_{\rm FD}(\R';\e,\e)=-\frac{2}{R'^{3}}+\frac{6(\boldsymbol{e}\cdot\boldsymbol{R}')^{2}}{R'^5},\\
        \boldsymbol{G}_{\rm RD}(\R';\e,\e)&=\frac{3(\boldsymbol{e}\times \boldsymbol{R}')(\boldsymbol{e}\cdot\boldsymbol{R}')}{R'^{5}}, &\quad & P_{\rm RD}(\boldsymbol{e},\boldsymbol{e};\boldsymbol{R}')=0.
    \end{align}
\end{subequations}
This formalism allows introducing the non-dimensional strength of the force and torque dipoles via $\alpha_{\rm FD}\equiv Fl/(8\pi a^{2}\mu\U)$ and $\alpha_{\rm RD}\equiv Ll/(8\pi a^{3}\mu\U)$, respectively. Thus, the velocity and pressure fields produced by a microswimmer are obtained as a sum of the force- and torque-dipole contributions: $\vec{u}'_{\rm free}=\boldsymbol{u}'_{\rm FD}+\boldsymbol{u}'_{\rm RD}$ and $p'_{\rm free} = p'_{\rm FD}+p'_{\rm RD}$.

To obtain the Green's function near a planar no-slip surface $\mathcal{S}_{0}$ with boundary condition $\boldsymbol{u}=\boldsymbol{0}$ on the surface, we use the method of images, which for a point-force has been outlined in Refs.~\cite{Blake_1971,Kim:2005}. The solution to Eq.~\eqref{stokeslet} near a planar rigid surface is then given by $\boldsymbol{u}'^{\rm I}_{\rm F}=(Fl/8\pi\mu a\U)\boldsymbol{G}^{\rm I}_{\rm F}$ where we note the expression of $\boldsymbol{G}^{\rm I}_{\rm F}$ for a tilted Stokeslet from Ref.~\cite{Spagnolie:2012}
\begin{align}\label{stokesletimage}
    \boldsymbol{G}^{\rm I}_{\rm F}(\boldsymbol{R}'^{\rm I};\e)=&\cos\theta\left(-\boldsymbol{G}_{\rm F}(\boldsymbol{R}'^{\rm I};\x')+2h \boldsymbol{G}_{\rm FD}(\boldsymbol{R}'^{\rm I};\x',\z')- 2h^{2}\boldsymbol{G}_{\rm D}(\boldsymbol{R}'^{\rm I};\x')\right)\nonumber\\&+\sin\theta\left(-\boldsymbol{G}_{\rm F}(\boldsymbol{R}'^{\rm I};\z')-2h \boldsymbol{G}_{\rm FD}(\boldsymbol{R}'^{\rm I};\z',\z')+2h^{2}\boldsymbol{G}_{\rm D}(\boldsymbol{R}'^{\rm I};\z')\right),
\end{align}
where $\boldsymbol{r}'^{\rm I}=(0,0,-h)$ is the image point and $\boldsymbol{G}_{\rm D}\equiv -(1/2)\nabla_{\boldsymbol{R}'^{\rm I}}^{2}\boldsymbol{G}_{\rm F}$ with $\boldsymbol{R}'^{\rm I}=\boldsymbol{r}'_{0}-\boldsymbol{r}'^{\rm I}$. The solution to the image pressure field is obtained by replacing the $\boldsymbol{G}_{(\cdot)}$ with the corresponding $P_{(\cdot)}$. Using Eq.~\eqref{stokesletimage}, we obtain the images for the Green's function of the force and torque dipoles, $\{\boldsymbol{G}^{\rm I}_{\rm FD},P^{\rm I}_{\rm FD}\}$ and $\{\boldsymbol{G}^{\rm I}_{\rm RD},P^{\rm I}_{\rm RD}\}$ respectively, by taking the derivatives of the image solution of the point force, in the same way, as was done above for the free space solutions. 

In our work, the microswimmer is modelled as a combination of a force and torque dipole with a non-dimensional free space propulsive velocity~$\e$. The planar-surface induced flow field solutions are given by $\boldsymbol{u}'^{\rm I}\equiv \alphafd \boldsymbol{G}^{\rm I}_{\rm FD}+\alphard \boldsymbol{G}^{\rm I}_{\rm RD}$ and $p'^{\rm I}\equiv \alphafd P^{\rm I}_{\rm FD}+\alphard P^{\rm I}_{\rm RD}$. The solution to the velocity and the pressure field of the zeroth-order problem is then obtained as $\boldsymbol{u}'^{(0)}=\boldsymbol{u}'_{\rm free}+\boldsymbol{u}'^{\rm I}$ and $p'^{(0)}=p'_{\rm free}+p'^{\rm I}$. 

We further note that the zeroth-order contributions to the translational and angular velocities follow from Faxen's law~\cite{Leal:2007,Spagnolie:2012}. In the lab frame of reference ($O'$) they evaluate to leading order to
\begin{subequations}	
 \begin{align}
\boldsymbol{U}^{(0)}&=\vec{e}+\boldsymbol{u}^{I}(\boldsymbol{r}_{0}) = \e+\alphafd\left[\frac{3}{8h^{2}}\sin2\theta{\vec{\hat{\rho}}_0}+\frac{3}{16h^{2}}(1-3\cos2\theta)\z'\right] ,\,\,\label{faxenvel}\\
\boldsymbol{\Omega}^{(0)}&=\frac{1}{2}\boldsymbol{\nabla} \times \boldsymbol{u}^{I}(\boldsymbol{r}_{0})=\frac{3\alphafd}{16h^{3}}\sin2\theta\vec{\hat{\phi}}_0+\alphard\left[\frac{9}{32h^{4}}\sin2\theta\vec{\hat{\rho}}_0+\frac{3}{64h^{4}}(1-3\cos 2\theta)\z'\right],\label{faxenangvel}
\end{align}
 \end{subequations}
with $\vec{\hat{\rho}}_0=\cos\varphi\x'+\sin \varphi\y'$ and ${\vec{\hat{\varphi}}}_0=\hat{\vec{z}}'\times {\vec{\hat{\rho}}_0}$. Here, (with respect to the lab frame of reference) the swimming direction is parameterized by the pitch angle $\theta$ and the polar angle $\varphi$, $\e=\cos\theta \vec{\hat{\rho}}_0+\sin \theta \z'$.

For convenience, in our work we move to the frame moving with the swimmer (with center at $O$ and coordinates $\{x,y,z\}$). The non-dimensional velocity and the pressure field in this frame are related to the fixed frame through $\boldsymbol{u}=\boldsymbol{u}'-\Uparand^{(0)}$ and $p'=p$ respectively, with the associated BCs in this frame, $\boldsymbol{u}=-\Uparand^{(0)}$ on $\mathcal{S}_{\infty}$ and $\mathcal{S}_{0}$. Note that this does not change the analysis, as the deformation [Eq.\,(1) of the main text] and the first-order deformation-induced velocities [in Eq.\,(3) of the main text] depend on the zeroth-order stress or the derivatives of the zeroth-order velocities, which are the same in both frames.

To mimic the near-field and steric interactions of the microswimmer with the boundary and to prevent numerical inaccuracies due to the swimmer physically colliding with the boundary, we follow previous work~\cite{Spagnolie:2012,Poddar:2020} by adding a short-ranged repulsive force of the form \begin{align}
    \boldsymbol{F}_{\rm rep}=A\exp(-Bh)/(1-\exp(-Bh))\z, \label{frepulsive}
\end{align} where $A$ and $B$ are suitably chosen such that the force is short-ranged and does not change the dynamics of the swimmer far from the boundary.
For the pusher we use $A=100$ and $B=2.8$ such that the closest distance of approach of the swimmer, when incident normal to the surface, is $h^\star\approx 1.5$. Similarly, for the puller swimmers we use $A=100$ and $B=2.5$, leading to $h^\star\approx1.5$.

\subsection{Calculation of the deformation \label{sec:deformation}} 
To compute the leading-order deformation $\delta^{(1)}$, the pressure produced by the swimming agent at the wall ($z=0$) is required as input
\begin{align} \label{pressurewallFD}
		p^{(0)}(z=0)=& \alphafd\left[-\frac{9h^{2}}{(\rho^{2}+h^{2})^{5/2}}+\frac{15h^{4}}{(\rho^{2}+h^{2})^{7/2}}+\frac{27 h^{2}\cos 2\theta}{(\rho^{2}+h^{2})^{5/2}}-\frac{45 h^{4}\cos 2\theta}{(\rho^{2}+h^{2})^{7/2}}+\frac{12h\rho\sin2\theta}{(\rho^{2}+h^{2})^{5/2}}\cos\phi \right. \nonumber\\&\qquad \qquad \left.-\frac{60h^{3}\rho\sin 2\theta}{(\rho^{2}+h^{2})^{7/2}}\cos\phi  +\frac{15h^{2}\rho^{2}}{(\rho^{2}+h^{2})^{7/2}}\cos 2\phi+\frac{15h^{2}\rho^{2}\cos 2\theta}{(\rho^{2}+h^{2})^{7/2}}\cos 2\phi \right]\\
	&+\alphard\left[-\frac{6\rho\sin 2\theta}{(\rho^{2}+h^{2})^{5/2}}\sin\phi+\frac{30h^{2}\rho\sin2\theta}{(\rho^{2}+h^{2})^{7/2}}\sin\phi-\frac{30h\rho^{2}\cos^{2}\theta}{(\rho^{2}+h^{2})^{7/2}}\sin 2\phi\right],\nonumber
\end{align}
where we introduced cylindrical coordinates $(z,\rho, \phi)$ with in-plane coordinates $\vec{r}_\parallel = (\rho \cos \phi, \rho \sin \phi)$. To obtain the Fourier transform  of Eq.~\eqref{pressurewallFD} as input for Eq.~\eqref{defft1}, we readily observe that it is expressible in the form of an angular mode decomposition: 
	\begin{align}
		p^{(0)}(\rho,\phi)=\sum_{m=-2}^{m=2} p^{(0)}_{m}(\rho) \exp(im\phi), \label{angledecomposition}
	\end{align}
with coefficients:
\begin{subequations}
	\begin{align}
		p^{(0)}_{-2}(\rho)&=\frac{\alphafd}{2}\left(\frac{15h^{2}\rho^{2}}{(h^2+\rho^2)^{7/2}}+\frac{15h^{2}\rho^{2}}{(h^2+\rho^2)^{7/2}}\cos 2\theta\right)+\frac{i\alphard}{2}\left(-\frac{30h\rho^{2}}{(h^2+\rho^2)^{7/2}}\cos^{2}\theta\right),\\
		p^{(0)}_{-1}(\rho)&=\frac{\alphafd}{2}\left(\frac{12h\rho}{(h^2+\rho^2)^{5/2}}-\frac{60h^{3}\rho}{(h^2+\rho^2)^{7/2}}\right)\sin 2\theta+\frac{i\alphard}{2}\left(-\frac{6\rho}{(h^2+\rho^2)^{5/2}}+\frac{30h^{2}\rho}{(h^2+\rho^2)^{7/2}}\right)\sin 2\theta,\\
		p^{(0)}_{0}(\rho)&=\alphafd\left(-\frac{9h^{2}}{(h^2+\rho^2)^{5/2}}+\frac{15h^{4}}{(h^2+\rho^2)^{7/2}}+\frac{27 h^{2}\cos 2\theta}{(h^2+\rho^2)^{5/2}}-\frac{45 h^{4}\cos 2\theta}{(h^2+\rho^2)^{7/2}}\right),\\
		p^{(0)}_{1}(\rho)&=\frac{\alphafd}{2}\left(\frac{12h\rho}{(h^2+\rho^2)^{5/2}}-\frac{60h^{3}\rho}{(h^2+\rho^2)^{7/2}}\right)\sin 2\theta-\frac{i\alphard}{2}\left(-\frac{6\rho}{(h^2+\rho^2)^{5/2}}+\frac{30h^{2}\rho}{(h^2+\rho^2)^{7/2}}\right)\sin 2\theta,\\
		p^{(0)}_{2}(\rho)&=\frac{\alphafd}{2}\left(\frac{15h^{2}\rho^{2}}{(h^2+\rho^2)^{7/2}}+\frac{15h^{2}\rho^{2}}{(h^2+\rho^2)^{7/2}}\cos 2\theta\right)-\frac{i\alphard}{2}\left(-\frac{30h\rho^{2}}{(h^2+\rho^2)^{7/2}}\cos^{2}\theta\right).
	\end{align}
\end{subequations}
This allows computing the two-dimensional Fourier transform ($\vec{r}_\parallel\to\vec{k} = k(\cos\phi_k, \sin\phi_k)$) in terms of the Hankel transform (see Sec.~\ref{sec:mathematics}) via
	 \begin{align}\label{pressure0Ft}
	p^{(0)}(k,\phi_{k})=2\pi\sum_{m=-2}^{2}i^{-m}p^{(0)}_{m}(k)\exp(im\phi_{k})\quad {\rm with}\quad p^{(0)}_{m}(k)=\int_{0}^{\infty} p^{(0)}_{m}J_{m}(k\rho) \rho \diff\rho,
	\end{align}
 where $J_{m}(\cdot)$ denotes the Bessel function of the first kind of $m-$th order. We obtain analytical expressions for $p_m^{(0)}(k)$ by following Ref.~\cite{bateman1954} in using the integral relation
\begin{align}
\int_0^\infty \frac{\rho^m}{(h^2+\rho^2)^{m+1/2}}J_m(k\rho)\rho \diff\rho &= \frac{\sqrt{\pi}k^{m-1}e^{-hk}}{2^m \Gamma(m+1/2)},
\end{align}
where $\Gamma(\cdot)$ denotes the Gamma function, and its derivatives with respect to $h$ as well as the relation between Bessel functions $J_{-m}(\cdot)= (-1)^mJ_m(\cdot)$. The latter are presented in Table~\ref{tab:p0ft}.
\begin{table}[htb]
		\begin{tabular}{|c|c|}
			\hline
			  $m$ & $\centering{p_{m}^{(0)}}(k)$\\ 
			 \hline
			$-2$ & $e^{-hk}k^{2}\left(h\alphafd-i\alpha_{RD}\right)\cos^{2}\theta$ \\ 
			\hline
			$-1$ & $-e^{-hk}k^{2}\left(-2h\alphafd+i\alpha_{RD}\right)\sin2\theta $ \\ 
			\hline
			$0$ & $ e^{-hk}hk^{2}(1-3\cos 2\theta)\alphafd$ \\ 
			\hline
			$1$ & $ e^{-hk}k^{2}\left(-2h\alphafd-i \alpha_{RD}\right)\sin 2\theta$ \\ 
			\hline
			$2$ & $ e^{-hk}k^{2}\left(h\alphafd+i\alpha_{RD}\right)\cos^{2}\theta$ \\ 
			\hline
		\end{tabular}
		\caption{Coefficients $p^{(0)}_m(k)$ of the Hankel transform of the pressure at the wall  $p^{(0)}(k,\phi)$ [Eq.~\eqref{pressure0Ft}]. }
		\label{tab:p0ft}
	\end{table}

 The Fourier transform of the deformation [Eq.~\eqref{defft1}] can then be expressed by 
	\begin{align}
		\delta^{(1)}(k,\phi_{k})&=2\pi\sum_{m=-2}^{2}i^{-m}\left(-\dfrac{1+\Gamma}{k^{4}+\Gamma k^{2}}p^{(0)}_{m}(k)\right)\exp(im\phi_{k}) \equiv \sum_{m=-2}^{2}\delta^{(1)}_{m}(k)\exp(im\phi_{k}), \label{deformationft}
	\end{align}
and derived in real space via an inverse Hankel transform (see Sec.~\ref{sec:mathematics}) via
	\begin{align}
		\delta^{(1)}(\rho,\phi)&=\frac{1}{2\pi}\sum_{m=-2}^{m=2}i^{m}\delta^{(1)}_{m}(\rho)\exp(im\phi)  \quad {\rm with}\quad \delta^{(1)}_{m}(\rho)=\int_{0}^{\infty} \delta^{(1)}_{m}(k)J_{m}(k\rho) k \diff k, \label{invdeffull}
	\end{align}
which we evaluate numerically. 

For some cases we can derive analytical predictions. In particular, the deformation at $\vec{r}_\parallel=\vec{0}$ can be computed from Eq.~\eqref{invdeffull}: 
\begin{align}
    \delta^{(1)}(\rho=0,\phi=0)=\frac{1}{2\pi}\sum_{m=-2}^{m=2}i^{m}\int_{0}^{\infty} \left\{-2\pi i^{-m}\dfrac{1+\Gamma}{k^{4}+\Gamma k^{2}}p^{(0)}_{m}(k)\right\} J_{m}(0) k \diff k. 
\end{align}
Since $J_{m}(0)=0$ for all $m\neq 0$ and $J_{0}(0)=1$, we have
\begin{align}
\delta^{(1)}(\rho=0,\phi=0)%&= \alphafd h (1-3\cos2\theta) \int_{0}^{\infty} \dfrac{1+\Gamma}{k^{4}+\Gamma k^{2}} \exp(-hk) k^{3} dk\\
&=\alphafd h (1-3\cos2\theta)\int_{0}^{\infty} \dfrac{1+\Gamma}{k^{2}+\Gamma} \exp(-hk) k\diff k.
\end{align}
Using Mathematica, we arrive at
\begin{align}
\delta^{(1)}(\rho=0,\phi=0)=\alphafd(1+\Gamma)h(1-3\cos2\theta)\left[\sin \left(\sqrt{\Gamma } h\right) \left(\frac{\pi}{2} - \text{Si}\left(h \sqrt{\Gamma }\right)\right)-\text{Ci}\left(h \sqrt{\Gamma }\right) \cos \left(\sqrt{\Gamma } h\right)\right], \label{eq:deformation0}
\end{align}
where the ${\rm Si}$ and the ${\rm Ci}$ are the Sine and Cosine intergals, respectively. Importantly, in the limit of small $\Gamma$, we observe a logarithmic divergence of the deformation: $\lim_{\Gamma \to 0} \delta^{(1)}(\rho=0,\phi=0) \sim \ln (h\sqrt{\Gamma})$. A similar logarithmic divergence has been observed in Refs.~\cite{Aderogba:1978,Berdan:1982}, which studied the deformation shape of a fluid-fluid interface due to a point force~\cite{Aderogba:1978} and a  sphere sedimenting towards the boundary~\cite{Berdan:1982}. This singularity has been shown to vanish for a swimmer moving towards a fluid-fluid interface~\cite{Shaik:2017}. In our work, however, we observe the logarithmic singularity for a swimmer near a deformable surface (endowed with surface tension and bending rigidity) for any swimming orientation, in the limit of $\Gamma\to 0$. We have chosen our parameters in a way that the deformation-induced velocities remain within the perturbation regime, yet for larger elasto-viscous numbers $\epsilon$ one has to carefully select the appropriate range of $\Gamma$ to account for the limitations of this membrane model. The logarithmic divergence of the deformation could be avoided by introducing a confining potential of the form $g$\,(in units of force per unit volume), such that the membrane equation\,[Eq.~\eqref{helfrich}] becomes modified to $(\kappa \tilde{\nabla}_{\parallel}^{4}-\Sigma\tilde{\nabla}_{\parallel}^{2}+g)\delta=\boldsymbol{\sigma}:\boldsymbol{n}\boldsymbol{n}$~\cite{Rallabandi:2018}. The confining potential $g$ may arise physically due to the finite size of the surface, confinement in an optical-trap, or the presence of the cytoskeleton in biological membranes. Upon non-dimensionalization and perturbative expansion, the first-order deformation is obtained from  $(\nabla_{\parallel}^{4}-\Gamma\nabla_{\parallel}^{2}+G)\delta^{(1)}=-(1+\Gamma+G)p^{(0)}$ with $G=ga^{4}/\kappa$.

\subsection{Calculation of the deformation-induced velocities \label{sec:velocities}}
\subsubsection{Computation of the velocities}
Using the first-order deformation $\delta^{(1)}$ as input, we compute the first-order correction to the swimming velocities, $\vec{U}^{(1)}$ and $\boldsymbol{\Omega}^{(1)}$, via Eq.~\eqref{reciprocal4}. We consider as auxiliary problem, the flow field due to a point force\,(or Stokeslet) and point torque of magnitude $\hat{F}_{i}(\hat{L}_{i}) = 8\pi$ along the $i{\rm th}$-direction, near a rigid-wall with no-slip boundary condition. The velocity fields of the auxiliary problem have been solved using the method of images~\cite{Blake_1971, Spagnolie:2012,Christina:2021}. Here,  we list the pressure and the normal components of the stresses on the wall\,($z=0$) due to a point force\,($\sigmahat^{{ F}, i}$) or a point torque \,($\sigmahat^{{ L},i}$) along the $i{\rm th}$-direction. Using cylindrical coordinates ($\rho,\phi,z$) and defining the vectors $\boldsymbol{\hat{\rho}}=\cos\phi\x+\sin\phi\y$ and $\boldsymbol{\hat{\phi}}=-\sin\phi\x+\cos\phi\y$ , we have  
\begin{subequations}
\label{stokesletstress}
\begin{align}
			\hat{p}^{F,x}&=\frac{12h^{2}\rho \cos\phi}{\left(h^2+\rho^2\right)^{5/2}}, &\quad 
			\sigmahat^{F,x}_{z}&= \frac{12 h \rho^2 \cos \phi}{\left(h^2+\rho^2\right)^{5/2}}\rhohat-\frac{12 h^2 \rho \cos \phi}{\left(h^2+\rho^2\right)^{5/2}}\z,\\ 
			\hat{p}^{F,y}&=\frac{12h^{2}\rho \sin\phi}{\left(h^2+\rho^2\right)^{5/2}},  &\quad 
			\sigmahat^{F,y}_{z}&=\frac{12 h \rho^2 \sin \phi}{\left(h^2+\rho^2\right)^{5/2}}\rhohat-\frac{12 h^2 \rho \sin \phi}{\left(h^2+\rho^2\right)^{5/2}}\z,\\ 
			\hat{p}^{F,z}& =-\frac{12h^{3}}{\left(h^2+\rho^2\right)^{5/2}}, &\quad
			\sigmahat^{F,z}_{z}& =-\frac{12 h^{2} \rho}{\left(h^2+\rho^2\right)^{5/2}}\rhohat+\frac{12 h^3}{\left(h^2+\rho^2\right)^{5/2}}\z, 
\end{align}
\end{subequations}
	and 
\begin{subequations}
\label{torquestress}
\begin{align}
			\hat{p}^{L,x}&=-\frac{12h\rho \sin\phi}{\left(h^2+\rho^2\right)^{5/2}}, &\quad   \sigmahat^{L,x}_{z}&=\frac{6 \left(h^2-\rho^2\right) \sin \phi}{\left(h^2+\rho^2\right)^{5/2}}\rhohat+\frac{6 h^2 \cos \phi}{\left(h^2+\rho^2\right)^{5/2}}\phihat+\frac{12 h \rho \sin \phi}{\left(h^2+\rho^2\right)^{5/2}}\z,\\
			\hat{p}^{L,y}&=\frac{12h\rho \cos\phi}{\left(h^2+\rho^2\right)^{5/2}}, &\quad
			\sigmahat^{L,y}_{z}&=\frac{6 \left(\rho^2-h^2\right) \cos \phi}{\left(h^2+\rho^2\right)^{5/2}}\rhohat+\frac{6 h^2 \sin \phi}{\left(h^2+\rho^2\right)^{5/2}}\phihat-\frac{12 h \rho \cos \phi}{\left(h^2+\rho^2\right)^{5/2}}\z, \\
			\hat{p}^{R,z}&=0, &\quad    \sigmahat^{L,z}_{z}&=\frac{6 h \rho}{\left(h^2+\rho^2\right)^{5/2}}\phihat. 
\end{align}
\end{subequations}
As last input for the reciprocal relation [Eq.~\eqref{reciprocal4}], we require 
\begin{align}\label{duz0}
		\left.\frac{\partial \ubold^{(0)}}{\partial z}\right|_{z=0}= \left.\frac{\partial \ubold^{(0)}_{\rm FD}}{\partial z}\right|_{z=0}+ \left.\frac{\partial \ubold^{(0)}_{\rm RD}}{\partial z}\right|_{z=0},
  \end{align}
with 
\begin{subequations}\label{duz0}
\begin{align}
\left.\frac{\partial \ubold_{\rm FD}^{(0)}}{\partial z}\right|_{z=0}=&\frac{3}{(\rho^{2}+h^{2})^{7/2}}\left[-2h\rho(h^{2}-4\rho^{2})\cos2\phi\cos^{2}\theta+h\rho(2\rho^{2}-3h^{2})(-1+3\cos2\theta)\right.\nonumber\\&\left.+2(h^{4}-8h^{2}\rho^{2}+\rho^{4})\cos\phi \sin2\theta \right]\rhohat+\frac{12h\sin\phi\cos\theta}{(\rho^{2}+h^{2})^{5/2}}\left[ \rho\cos\phi\cos\theta-h\sin\theta\right]\phihat\\
\left.\frac{\partial \ubold_{\rm RD}^{(0)}}{\partial z}\right|_{z=0}=&\frac{12\cos\theta\sin\phi}{(h^{2}+\rho^{2})^{7/2}}\left[\rho(3h^{2}-2\rho^{2})\cos\phi\cos\theta+h(3\rho^{2}-2h^{2})\sin\theta\right]\rhohat +\frac{3}{2(\rho^{2}+h^{2})^{7/2}}\Bigl[2\rho(6h^{2}+\rho^{2})\notag\\
&\times\cos2\phi\cos^{2}\theta-\rho(\rho^{2}-4h^{2})(3\cos2\theta-1)+4h(3\rho^{2}-2h^{2})\cos\phi\sin 2\theta\Bigr]\phihat
\end{align}
\end{subequations}
We further use the zeroth-order translational velocity along $\vec{\rho}_0$
	\begin{align}\label{U0parallel}
		U^{(0)}_{\parallel}=\cos \theta+\alphafd\frac{3}{8h^{2}}\sin2\theta,
	\end{align}	
and that
	\begin{align}
		\left.\frac{\partial \hat{u}^{i}_{z}}{\partial z}\right|_{z=0}=0 \quad{\rm and}\quad   \left.\frac{\partial u^{(0)}_{z}}{\partial z}\right|_{z=0}=0.
	\end{align}

To compute the integral in Eq.~\eqref{reciprocal4}, rather than working in real space, we  exploit Parseval's identity
 \begin{align}
		\int_{\mathcal{R}^{2}}\,f(\boldsymbol{r})g(\boldsymbol{r}) \ {\diff}^{2}r=\frac{1}{(2\pi)^{2}}\int_{\mathcal{R}^{2}}\,f(\boldsymbol{k})[g(\boldsymbol{k})]^* \  \diff^2 k, \label{parseval}
\end{align}
where $[\cdot]^*$ denotes the complex conjugate. We further assume an instantaneous boundary deformation, compared to the swimming time scale, $\partial_t\delta\approx 0$. 
Thus, the integral in Eq.~\eqref{reciprocal4} can be reformulated as 
\begin{subequations}
  \begin{align}
U^{(1)}_{i}\,(	\Omega^{(1)}_{i})&=\frac{1}{8\pi}\int \,\left[\hat{p}^{(i)}_{F\, (L)}U_{\parallel}^{(0)}\frac{\partial}{\partial x}- \boldsymbol{\hat{z}}\cdot \boldsymbol{\hat{\sigma}}_{F\, (L)}^{(i)}\cdot\frac{\partial \boldsymbol{u}^{(0)}}{\partial z}\right]_{z=0}\delta^{(1)} \ \diff^2 r \label{app:vel_real} \\
&=\frac{1}{(2\pi)^{2}8\pi}\int_{0}^{\infty}\int_{0}^{2\pi} \,\left[U_{\parallel}^{(0)}\hat{p}^{(i)}_{F\, (L)}\left[ik\cos\phi_{k}\delta^{(1)}\right]^{*}-\boldsymbol{\hat{z}}\cdot \boldsymbol{\hat{\sigma}}_{F\, (L)}^{(i)}\cdot\frac{\partial \boldsymbol{u}^{(0)}}{\partial z}\left[\delta^{(1)}\right]^{*}\right]_{z=0} \ \diff\phi_k k\diff k \\
&=-\frac{1}{32\pi^{3}}\int_{0}^{\infty}\int_{0}^{2\pi}\,\left[ikU_{\parallel}^{(0)}\hat{p}^{(i)}_{F\, (L)}\cos\phi_{k}+\boldsymbol{\hat{z}}\cdot \boldsymbol{\hat{\sigma}}_{F\, (L)}^{(i)}\cdot\frac{\partial \boldsymbol{u}^{(0)}}{\partial z}\right]_{z=0}\left[\delta^{(1)}\right]^{*} \ \diff\phi_k k\diff k,\label{parsevalft}
\end{align}  
\end{subequations}
where $\hat{p}^{(i)}_{{F}\,({L})}$ and $\boldsymbol{\hat{\sigma}}_{F\,(L)}^{(i)}$ are the auxiliary pressure and stress due to the force\,($F$)\,(or torque ($L$)) along the $i^{\rm th}$-direction, respectively. We further decompose the integrands in their angular modes, 
 \begin{subequations}
\begin{align}
  ikU_{\parallel}^{(0)}\cos\phi_{k}\hat{p}^{(i)}_{{F}\,({L})} &= \sum_{m} G_{m, { F({ L})}}^{(i)}(k) \exp(i m\phi_k),\label{angledecompositionpressure}\\ 
\boldsymbol{\hat{z}}\cdot \boldsymbol{\hat{\sigma}}^{(i)}_{F(L)}\cdot\left.\frac{\partial \boldsymbol{u}^{(0)}}{\partial z}\right|_{z=0}&= \sum_m F_{m, {F\,(L)}}^{(i)}(k)\exp(i m\phi_k), \label{angledecompositionsigma}
\end{align}
\end{subequations}
with coefficients $G_{m, { F\,(L)}}^{(i)}(k)$ and $F_{m, {F\,(L)}}^{(i)}(k)$, which we will derive in what follows.

For the first integrand, we use the Hankel transforms of the pressure of the auxiliary problem, 
\begin{subequations}
 \begin{align}
		\hat{p}^{x}_{F}&=-8\pi i hk\exp(-hk)\cos\phi_{k}, & \quad \hat{p}^{y}_{F}&=-8\pi ihk\exp(-hk)\sin\phi_{k}, \\
		\hat{p}^{z}_{F}&=-8\pi(1+hk)\exp(-hk), & \quad \hat{p}^{x}_{L}&=8\pi ik\exp(-hk)\sin\phi_{k}, \\
		\hat{p}^{y}_{L}&=-8\pi ik\exp(-hk)\cos\phi_{k}, &\quad \hat{p}^{z}_{L}&=0, 
	\end{align}
 \end{subequations}
as input for the expansion in Eq.~\eqref{angledecompositionpressure}, whose coefficients are provided in Tab.~\ref{tab:advection}. 

\begin{table}[htp]
		\begin{tabular}{|p{0.5cm}|>{\raggedright\arraybackslash}p{2.5cm}|>{\raggedright\arraybackslash}p{2.75cm}|>{\raggedright\arraybackslash}p{2.5cm}|>{\raggedright\arraybackslash}p{2.75cm}|>{\raggedright\arraybackslash}p{2.5cm}|}
			\hline
			 & $m=-2$ & $m=-1$& $m=0$ & $m=1$&$m=2$\\ 
			 \hline
			$G^{x}_{F}$ & $2\pi U_{\parallel}^{(0)}hk^{2}e^{-hk}$ & $0$& $4\pi U_{\parallel}^{(0)}hk^{2}e^{-hk}$ & $0$&$2\pi U_{\parallel}^{(0)}hk^{2}e^{-hk}$\\ 
			\hline
			$G^{y}_{F}$ & $2\pi i U_{\parallel}^{(0)}hk^{2}e^{-hk}$ & $0$& $0$ & $0$&$-2\pi i U_{\parallel}^{(0)}hk^{2}e^{-hk}$\\ 
			\hline
			$G^{z}_{ F}$ & $0$ & $-4\pi iU_{\parallel}^{(0)}(1+hk)ke^{-hk}$& $0$ & $-4\pi iU_{\parallel}^{(0)}(1+hk)ke^{-hk}$&$0$\\ 
			\hline
			$G^{x}_{ L}$ & $-2\pi i U_{\parallel}^{(0)}k^{2}e^{-hk}$ & $0$& $0$ & $0$&$2\pi i  U_{\parallel}^{(0)}k^{2}e^{-hk}$\\ 
			\hline
			$G^{y}_{ L}$ & $2\pi U_{\parallel}^{(0)}k^{2}e^{-hk}$ & $0$& $4\pi U_{\parallel}^{(0)}k^{2}e^{-hk}$ & $0$&$2\pi U_{\parallel}^{(0)}k^{2}e^{-hk}$\\ 
			\hline$G^{z}_{L}$ & $0$ & $0$& $0$ & $0$&$0$\\ 
			\hline
		\end{tabular}
		\caption{Coefficients of the expansion in Eq.~\eqref{angledecompositionpressure}. }
		\label{tab:advection}
	\end{table}	

		\begin{table}[]
			\begin{tabular}{|p{0.5cm}|p{0.5cm}|>{\raggedright\arraybackslash}p{4.5cm}|>{\raggedright\arraybackslash}p{4.5cm}|>{\raggedright\arraybackslash}p{4.5cm}|}
				\hline
				$m$ & $n$ & $A^{x,{F}}_{m,n}$ & $A^{y,{F}}_{m,n}$ & $A^{z,{F}}_{m,n}$\\ \hline
				$0$ & $2$ & $36\alphafd h\sin 2\theta$ & $0$ & $0$ \\ \cline{2-5}
				& $3$ & $-396\alphafd h^{3}\sin 2\theta$ & $108\sin2\theta h^{2}\alphard$ & $\alphafd(72-216\cos2\theta)h^{3}$ \\\cline{2-5}
				& $4$ & $720\alphafd h^{5}\sin 2\theta$ & $-288\sin2\theta h^{4}\alphard$ & $\alphafd(-252+756\cos2\theta)h^{5}$ \\\cline{2-5}
				& $5$ & $-360\alphafd h^{7}\sin 2\theta$ & $180\sin2\theta h^{6}\alphard$ & $(180-540\cos2\theta)h^{7}\alphafd$ \\\hline
				
				$1$ & $3$ & $144\alphafd h^{2}\cos 2\theta+i36\alphard h\cos^{2}\theta$ & $i144\alphafd h^{2}\sin^{2}\theta-36\alphard h\cos^{2}\theta$ & $-36\alphafd h^{2}\sin 2\theta$\\\cline{2-5}
				& $4$ & $9h^{4}(5-51\cos2\theta)\alphafd-i126\alphard{\alpha} h^{3}\cos^{2}\theta$  & $-i\alphafd 9h^{4} (23-33\cos 2\theta)+126 h^{3} \cos^{2}\theta \alphard$ & $\alphafd 360 h^{4}\sin 2\theta+i108\alphard h^{3}\sin 2\theta$ \\ \cline{2-5}
				& $5$ & $\alphafd 45h^{6}(-1+7\cos2\theta)+i\alphard 90 h^{5}\cos^{2}\theta$ & $-i\alphafd 45h^{6}(-3+5\cos 2\theta)-\alphard 90 h^{5}\cos^{2}\theta$ & $-\alphafd 360 h^{6}\sin 2\theta -i\alphard 180 h^{5} \sin 2\theta$ \\\hline
				
				$2$ & $3$ & $\alphafd 18h \sin2\theta$ & $-i\alphafd 18 h \sin 2\theta$ & $0$\\  \cline{2-5}
				& $4$ & $-\alphafd 180h^{3}\sin2\theta-i\alphard 54 h^{2}\sin2\theta$ & $i\alphafd 180 h^{3} \sin 2\theta-\alphard 54 h^{2}\sin 2\theta$ & $-\alphafd 144 h^{3}\cos^{2}\theta+i \alphard 72 h^{2} \cos^{2}\theta$ \\  \cline{2-5}
				& $5$ & $\alphafd 180h^{5}\sin2\theta+i\alphard90\tilde h^{4}\sin2\theta$ & $-i\alphafd 180 h^{5}\sin 2\theta+\alphard 90 h^{4} \sin 2\theta$ & $\alphafd 180 h^{5} \cos^{2}\theta+i180 \alphard h^{4}\cos^{2}\theta $ \\ \hline

				$3$ & $4$ & $\alphafd 72h^{2} \cos^{2}\theta+i\alphard 36h\cos^{2}\theta$ & $-i \alphafd 72 h^{2}\cos^{2}\theta+\alphard 36 h \cos^{2}\theta$ & $0$ \\ \cline{2-5}
				& $5$ & $-\alphafd 90h^{4}\cos^{2}\theta-i90\alphard h^{3}\cos^{2}\theta$ & $i\alphafd 90 \cos^{2}\theta h^{4}-\alphard 90 h^{3}\cos^{2}\theta$ & $0$ \\\hline\hline
				$m$ & $n$ & $A^{x,{L}}_{m,n}$& $A^{y,{L}}_{m,n}$ & $A^{z,{L}}_{m,n}$\\ \hline
				$0$ & $2$ & $0$ & $\alphafd 18 \sin 2\theta$ & $0$ \\ \cline{2-5}
				& $3$ & $-\alphard 54 h\sin 2\theta$ & $-\alphafd 216h^{2}\sin 2\theta$ & $\alphard h(9-27\cos 2\theta)$ \\\cline{2-5}
				& $4$ & $\alphard 252 h^{3} \sin 2\theta$ & $\alphafd 522 h^{4} \sin 2\theta$ & $\alphard h^{3}(-54+162 \cos 2\theta)$ \\\cline{2-5}
				& $5$ & $-\alphard 270 h^{5} \sin 2\theta$ & $-\alphafd 360 h^{6} \sin 2\theta$ & $\alphard h^{5} (45-135 \cos 2\theta)$ \\\hline
				
				$1$ & $3$ & $-i \alphafd 72 h \sin^{2}\theta+\alphard 18 \cos^{2}\theta$ & $\alphafd 72 h \cos 2\theta+i\alphard 18 \cos^{2}\theta$ & $0$\\\cline{2-5}
				& $4$ & $-i\alphafd 27 h^{3}(-5+7\cos 2\theta)-\alphard \frac{9}{4} h^{2}(15+23\cos 2\theta)$  & $\alphafd 27 h^{3}(1-11\cos 2\theta)-i \alphard \frac{9}{4} h^{2}(19+11 \cos 2\theta)$ & $i\alphafd 18h^{3} \sin 2\theta +\alphafd 108 h^{2}\sin 2\theta $ \\ \cline{2-5}
				& $5$ & $-i\alphafd 45 h^{5}(3-5\cos 2\theta)+\alphard \frac{45}{4}h^{4}(3+11\cos 2\theta)$ & $\alphafd 45h^{5}(-1+7\cos2\theta)-i\alphard \frac{45}{4}h^{4}(-7+\cos 2\theta) $ & $-\alphard 90 h^{4}\sin 2\theta$ \\\hline
				
				$2$ & $3$ & $i\alphafd 9 \sin 2\theta$ & $\alphafd 9 \sin 2\theta$ & $0$\\  \cline{2-5}
				& $4$ & $-i\alphafd 99 h^{2}\sin 2\theta+\alphard 27 h \sin 2\theta$ & $-\alphafd 99 h^{2}\sin 2\theta-i \alphard 27 h \sin 2\theta$ & $-i\alphafd 18 h^{2}\cos^{2}\theta+\alphard 9 h\cos^{2}\theta$ \\  \cline{2-5}
				& $5$ & $i\alphafd 180 h^{4}\sin 2\theta-\alphard 45 h^{3} \sin 2\theta$ & $\alphafd 180 \sin 2\theta h^{4}+\alphard 45 h^{3}\sin 2\theta$ & $\alphard 45 h^{3}\cos^{2}\theta$ \\ \hline

				$3$ & $4$ & $i\alphafd 36 h \cos^{2}\theta-\alphard 18 \cos^{2}\theta$ & $\alphafd 36 h \cos^{2}\theta+i\alphard 18 \cos^{2}\theta$ & $0$ \\ \cline{2-5}
				& $5$ & $-i\alphafd 90 h^{3}\cos^{2}\theta+\alphard \frac{135}{2} h^{3}\cos^{2}\theta$ & $-\alphafd 90 h^{3}\cos^{2}\theta-i\alphard\frac{135}{2} h^{2} \cos^{2}\theta$ & $0$ \\\hline
			\end{tabular}
			\caption{Coefficients of the expansion in Eqs.~\eqref{zsigmadudz0exp}-\eqref{zsigmadudz0ftexp}.}
			\label{tab:Amn}
		\end{table}

The Fourier transform of the second integrand [Eq.~\eqref{angledecompositionsigma}] is more tedious to compute. First, we express the integrand in real space ($\vec{r}_\parallel$) in terms of an angular mode decomposition. We define the coefficients $ A^{(i),{F}({L})}_{m,n}$,  where $m$ denotes the order of the angular part $\exp(im\phi)$, $n$ denotes the power of $(h^{2}+\rho^{2})^{n+1}$ in the denominator, $i\in\{x,y,z\}$ denotes the direction along which the point force (${F}$) or torque (${L}$) of the auxiliary problem is oriented. 
Using Eqs.~\eqref{stokesletstress},\eqref{torquestress} and \eqref{duz0} we obtain
\begin{align}
		F_{F\,(L)}^{(i)}(\rho,\phi)=\sum_{m=-3}^{m=3}\sum_{n} \frac{\rho^{m}A_{m,n}^{(i),{F}\,({L})}}{\Rbar^{n+1}}\exp(im\phi),     \label{zsigmadudz0exp}
\end{align}
 where the coefficients $A_{m,n}^{(i),{F}\,({L})}$ are listed in Tab.~\ref{tab:Amn}. Note that the values of $A_{m,n}^{(i),{F}\,({L})}$ are listed only for $m\geq 0$ as $A_{-m,n}^{(i),{F}\,({L})}=\left[A_{m,n}^{(i),{F}\,({L})}\right]^{*}$.
A Hankel transform [Eq.~\eqref{eq:Hankel} in Sec.~\ref{sec:mathematics}] of Eq.~\eqref{zsigmadudz0exp} yields the integrand in Fourier space
	\begin{align}
		F_{F\,(L)}^{(i)}(k,\phi_{k}) = 2\pi \sum_{m=-3}^{m=3} i^{-m}\sum_{n} A_{m,n}^{(i),{F}\,({L})}B_{n}^{m}\equiv\sum_{m=-3}^{m=3} F_{m,F\,(L)}^{(i)}(k)\exp(im\phi_{k}),\label{zsigmadudz0ftexp}
	\end{align}
where we have abbreviated~\cite{bateman1954} 
\begin{align}
		\int_{0}^{\infty}\frac{\rho^{m}}{(\rho^{2}+h^{2})^{n+1}}\,J_{m}(\rho k)\,\rho\, \diff \rho=\frac{h^{m-n}k^{n}K_{m-n}(hk)}{2^{n}\Gamma(n+1)}\equiv B_{n}^{m}. \label{ft2}
	\end{align}
Here, $K_{m}(\cdot)$ is the Bessel function of the second kind  and $\Gamma(\cdot)$ is the Gamma function. Additionally, we note that  $B_{n}^{-m}=(-1)^{m}B_{n}^{m}$, resulting from the property of the Bessel functions $J_{-m}(x)=(-1)^{m}J_{m}(x)$.

Finally, inserting the angular mode decomposition [Eqs.~\eqref{angledecompositionpressure}-\eqref{angledecompositionsigma}] into Eq.~\eqref{parsevalft} and noting that $\int_0^{2\pi}\exp(i (m-\ell)\phi_k)\diff\phi_k= 2\pi\delta_{\ell m}$, we arrive at 
\begin{align}
U^{(1)}_{i}\,(\Omega^{(1)}_{i})&=-\frac{1}{16\pi^{2}} \sum_{m=-2}^{2} \int_{0}^{\infty}\left[G_{m,{ F}\,({ L})}^{(i)}(k)+F_{m,{ F}\,({L})}^{(i)}(k)\right] [\delta_{m}(k)]^{*}\ k\diff k, \label{firstordervel}
\end{align}
which can be efficiently evaluated numerically.

\subsubsection{Numerical validation}
\begin{figure}[htp]
\centering
\includegraphics[width=\columnwidth]{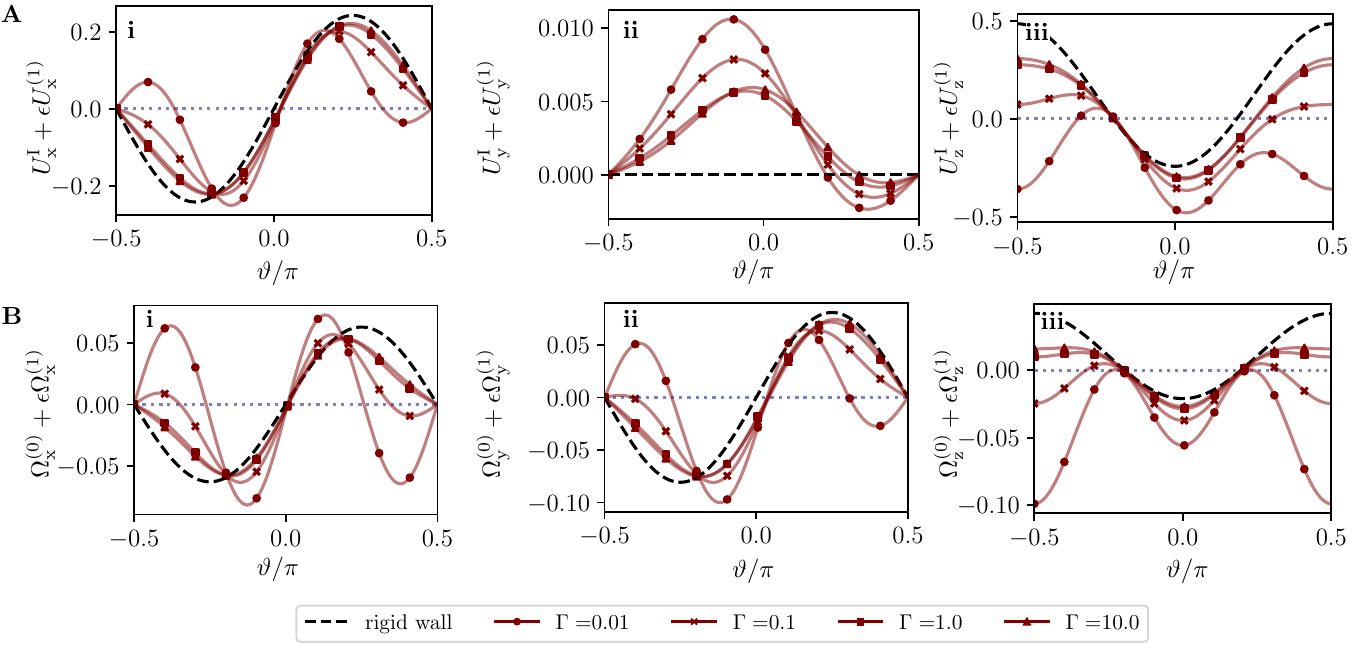}
\caption{(\textbf{A}) Sum of  the image-induced and the first-order velocity along the (\textbf{i})~$x-$ (\textbf{ii})~$y-$ and (\textbf{iii})~$z-$ direction respectively, and (\textbf{B}) deformation induced angular velocity\,($\Omega_{i}$) along the (\textbf{i})~$x-$,  (\textbf{ii})~$y-$, and (\textbf{iii})~$z-$direction respectively, for a pusher and different values of the FvK number\,($\Gamma$). The solid lines and markers are obtained by numerically integrating Eq.~\eqref{parsevalft} and Eq.~\eqref{app:vel_real}, respectively. The parameters for the pusher are the same as in the main text and $\epsilon=0.1 $}\label{fig:validation}
\end{figure}

To validate our velocities obtained by the method of Hankel transforms, we evaluate Eq.~\eqref{app:vel_real} in real space by computing the deformation $\delta^{(1)}$ numerically using Eq.~\eqref{invdeffull}. It is important to note that a direct numerical evaluation is much more costly, yet to assure our analytical calculations are correct we compare them for a selected set of parameters in Fig.~\ref{fig:validation}, showing excellent agreement between the two approaches.

%%%%%%%%%%%%%%%%%%%%%%%%%%%%%%%%%%%%%%%%%%%%%%%%%%%%%%%%%%%
\section{Additional mathematical details\label{sec:mathematics}}
Here, we derive the relation between the two-dimensional Fourier transform and the Hankel transform. We use polar coordinates in two dimensions in real and in Fourier space,  $\vec{r}_{\parallel}=(\rho \cos \phi,\rho \sin \phi)$ and $\vec{k}=(k\cos \phi_k,k\sin \phi_k)$, respectively. The Fourier transform of a function $f(\boldsymbol{r}_{\parallel})=f(\rho,\phi)$ can be expressed as 
\begin{align}
    f(k,\phi_{k})=\int_{0}^{\infty} \int_{0}^{2\pi} e^{-i\boldsymbol{k}\cdot\boldsymbol{r}_{\parallel}}f(\rho,\phi) \ \diff \phi \rho \diff \rho =\int_{0}^{\infty} \int_{0}^{2\pi}  e^{-ik\rho \cos(\phi-\phi_{k})}f(\rho,\phi) \ \diff \phi \rho \diff \rho . \label{ft2d}
\end{align}
Employing the Jacobi-Anger relation we can express~\cite{Abramowitz:1974}
\begin{align}
    e^{-ik\rho \cos(\phi-\phi_{k})}=\sum_{m=-\infty}^{\infty} i^{m} J_{m}(-k\rho)e^{im(\phi-\phi_{k})}=\sum_{m=-\infty}^{\infty} (-i)^{m} J_{m}(k\rho)e^{im(\phi-\phi_{k})},
\end{align}
where we have used that $J_{m}(-x)=(-1)^{m}J_{m}(x)$.
For functions $f(\rho,\phi)$ expressible in terms of an angular mode decomposition
\begin{align}
    f(\rho,\phi)=\sum_{n=-\infty}^{\infty}f_{n}(\rho) \exp(in\phi),
\end{align}
Eq.~\eqref{ft2d} can be rewritten as
\begin{align}
    f(k,\phi_{k})=\int_{0}^{\infty}\int_{0}^{2\pi} \sum_{m=-\infty}^{\infty} i^{-m} J_{m}(k\rho)\exp(im(\phi-\phi_{k}))\sum_{n=-\infty}^{\infty}f_{n}(\rho) \exp(in\phi) \ \diff \phi \rho \diff \rho\label{eq:JA}
\end{align}
Using that $\int_{0}^{2\pi}\exp(i\phi(m+n)) \diff\phi=2\pi\delta_{m,-n}$ and replacing $m$ by $-m$, Eq.~\eqref{eq:JA} simplifies to 
\begin{align}
     f(k,\phi_{k})=2\pi\sum_{m=-\infty}^{\infty} i^{-m} f_m(k)\exp(im\phi_{k}) \quad {\rm with} \quad f_m(k)=\int_{0}^{\infty} f_{m}(\rho)J_{m}(k\rho) \  \rho \diff \rho. \label{eq:Hankel}
\end{align}
The inverse Fourier transform is defined as
\begin{align}
   f(\rho,\phi)=\frac{1}{(2\pi)^2}\int_{0}^{\infty}\int_{0}^{2\pi} e^{i\boldsymbol{k}\cdot\boldsymbol{r}_\parallel}f(k,\phi_{k}) \ \diff \phi_k k \diff k,\label{ift2}
\end{align}
where the factor $(2\pi)^{-2}$ results from normalization. Following the same steps as above, we arrive at the inverse Hankel transform: 
\begin{align}
      f(\rho,\phi)=\frac{1}{2\pi}\sum_{m=-\infty}^{\infty} i^{m} f_{m}(\rho) \exp(im\phi) \quad {\rm with} \quad  f_{m}(\rho)=\int_{0}^{\infty} f_{m}(k) J_{m}(k\rho) \ k \diff k. \label{eq:inverseHankel}
\end{align}

\section{Additional results}
To understand the dependency of the phase plots and our observations on:\,elasto-viscous number\,$\epsilon$, the short-ranged repulsive force $\vec{F}_{\rm rep}$\,(see Eq.~\eqref{frepulsive}), and the initial height\,$h(0)$, we plot the phase diagrams by varying these parameters for the particular case of the pusher swimmer. 

\subsection{Variation of the elasto-viscous number}
Here, we show the phase plots of a pusher for varying elasto-viscous number $\epsilon$. The results for the planar rigid wall are recovered for a very small elasto-viscous number $\epsilon=0.01$. In the circular phase\,($\vartheta^{\star}=0$)[Fig.~\ref{fig:pusher_force}\textbf{B}] the pusher is aligned parallel to the wall and moves along circular trajetories.

\begin{figure}[htp]
\centering
{\includegraphics[width = 0.8\columnwidth]{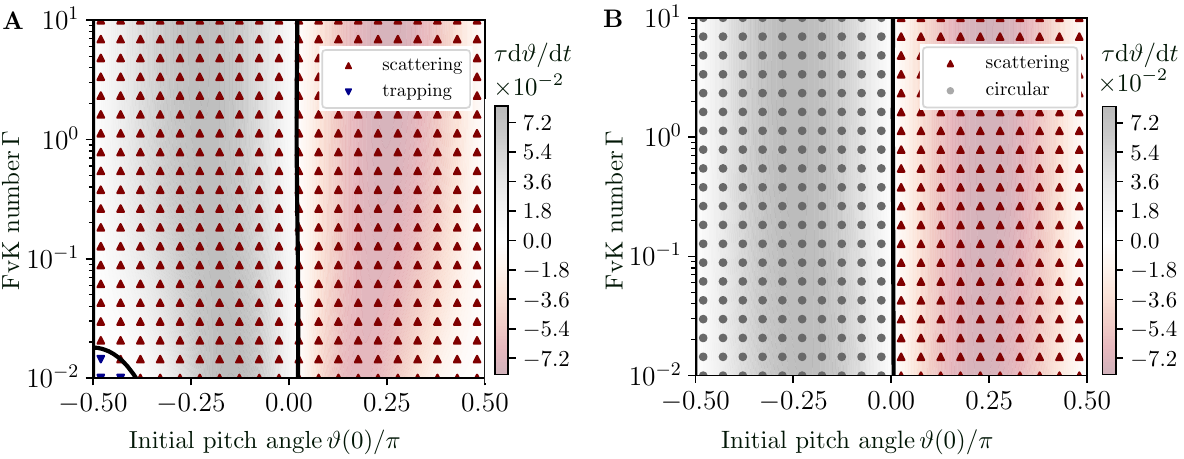}}

\caption{Phase diagram for a pusher (parameters as described in the main text) as a function of the initial orientation~$\vartheta(0)$ and Föppl-von K{\'a}rm{\'a}n number~$\Gamma$ for different elasto-viscous numbers: ({\bf A}) $\epsilon = 0.05$ and ({\bf B}) $\epsilon = 0.01$. Background colors indicate $\diff\vartheta/\diff t$ and black lines correspond to the associated fixed points. In ({\bf B}) circles indicate phases of circular near-surface motion. }\label{fig:pusher_epsilon}
\end{figure}

\subsection{Variation of the repulsive force}
\begin{figure}[htp]
\centering
{\includegraphics[width = 0.8\columnwidth]{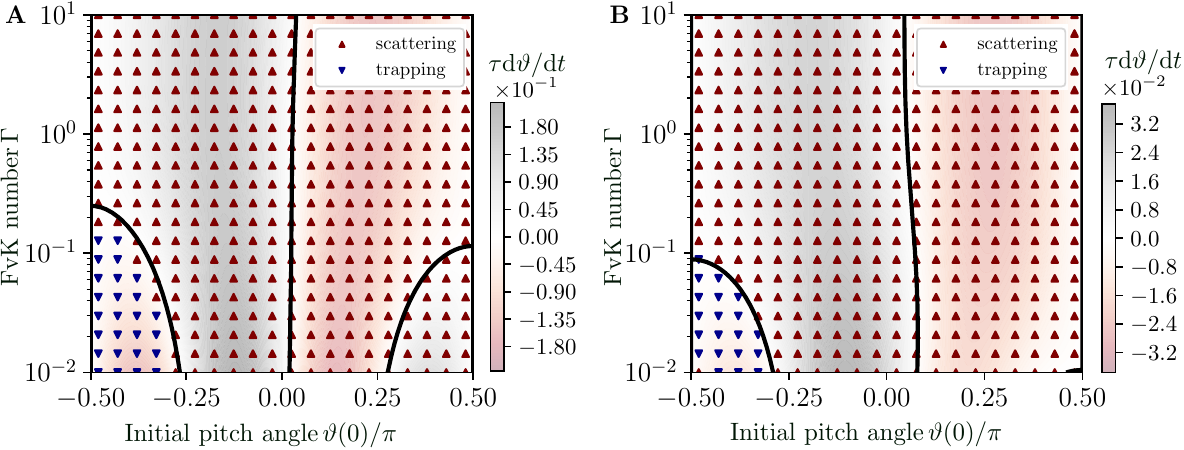}}

\caption{Phase diagram for a pusher (parameters as described in the main text) as a function of the parameter~$\vartheta(0)$ and Föppl-von K{\'a}rm{\'a}n number~$\Gamma$ for two different values of $B$ in $\vec{F}_{\rm rep}$: ({\bf A}) $B= 3.5$ and ({\bf B}) $B = 2.0$. Background colors indicate $\diff\vartheta/\diff t$ at $h^{\star}$ and black lines correspond to the associated fixed points.}\label{fig:pusher_force}
\end{figure}

Here, we plot the phase diagram by changing the repulsive force $\vec{F}_{\rm rep}$ in Eq.~\eqref{frepulsive}, which mimics the near-field and steric interactions of the swimmer with the boundary. In Fig.~\ref{fig:pusher_force}\textbf{A} we use $B=3.5$ while in Fig.~\ref{fig:pusher_force}\textbf{B} we set $B=2$, which results in a minimum distance of approach of $h^{\star}\approx 1.15$  and $h^{\star}\approx 2.25$, respectively. The phase behavior obtained from the trajectory calculations qualitatively remains the same. The swimmer exhibits the scattering and trapping state as discussed in the main text. 

\subsection{Variation of the initial height }
\begin{figure}[htp]
\centering
{\includegraphics[width = 0.8\columnwidth]{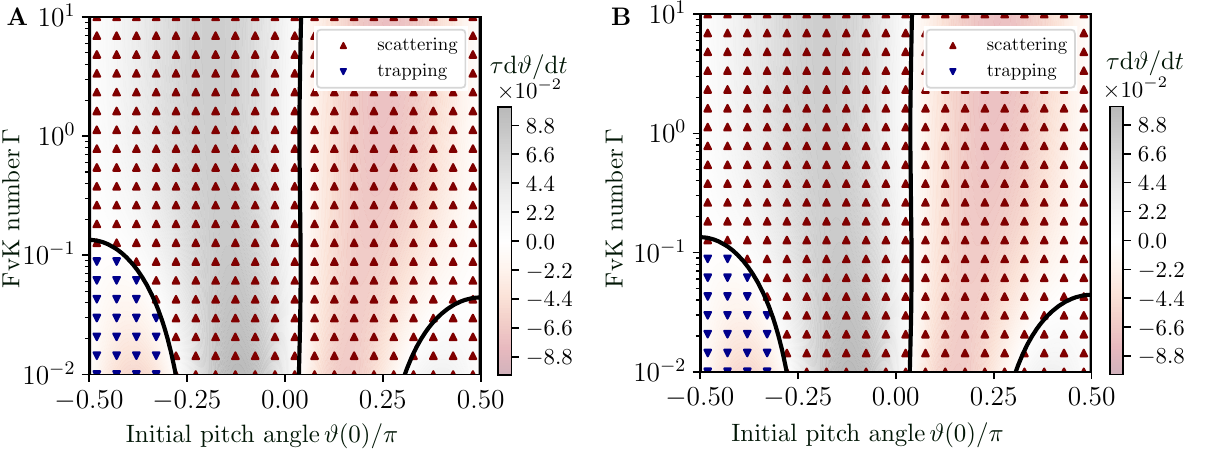}}

\caption{Phase diagram for a pusher (parameters as described in the main text) as a function of the parameter~$\vartheta(0)$ and Föppl-von K{\'a}rm{\'a}n number~$\Gamma$ for two different values of $h(0)$: ({\bf A}) $h(0)= 2a$ and ({\bf B}) $h(0) = 6a$. Background colors indicate $\diff\vartheta/\diff t$ at $h^{\star}$ and black lines correspond to the associated fixed points.}\label{fig:pusher_h}
\end{figure}

Here we plot the phase diagram by changing the initial height $h(0)$ for the trajectory calculation. All other parameters are the same as in the main text. In Fig.~\ref{fig:pusher_h}\textbf{A} and Fig.~\ref{fig:pusher_h}\textbf{B} we start from $h(0)=2a$ and $h(0)=6a$, respectively. The minimum distance of approach\,($h^{\star}\approx 1.5$) is the same as in the main text. There is no difference in the obtained phases and the value of $\diff \vartheta/\diff t$ represented by the background colors. This confirms that the phase plot is unaffected by the choice of the initial position. 

\bibliography{literature}